\documentclass[aps,prd,reprint,twocolumn,10pt,eqsecnum,showpacs,nofootinbib,amsfonts,amssymb,amsmath,longbibliography,superscriptaddress,floatfix]{revtex4-1}
\usepackage[colorlinks=true,allcolors=blue]{hyperref}
\usepackage{bm}
\usepackage{xcolor}
\usepackage{mathtools}
\usepackage{mathrsfs}
\usepackage{graphicx}
\usepackage{enumerate}

\DeclareSymbolFont{bbold}{U}{bbold}{m}{n}
\DeclareSymbolFontAlphabet{\mathbbold}{bbold}

\newcommand{\ud}{\mathrm{d}}
\newcommand{\uD}{\mathrm{D}}
\newcommand{\pb}[1]{\,\mbox{}_{#1}}

\newcommand{\ord}[2]{\underset{^{(#1)}}{#2}{}}
\newcommand{\bs}[1]{\boldsymbol{#1}}
\newcommand{\ms}[1]{\mathscr{#1}}
\newcommand{\mc}[1]{\mathcal{#1}}

\DeclareMathOperator*{\STF}{STF}

\usepackage[normalem]{ulem}

\begin{document}

\author{Alexander M.\ Grant}
\email{alex.grant@virginia.edu}

\author{David A.\ Nichols}
\email{david.nichols@virginia.edu}

\affiliation{Department of Physics, University of Virginia, P.O.~Box 400714,
  Charlottesville, Virginia 22904-4714, USA}

\title{Persistent gravitational wave observables: \\ Curve deviation in asymptotically flat spacetimes}

\begin{abstract}
  In the first paper in this series, a class of observables that generalized the gravitational wave memory effect were introduced and given the name ``persistent gravitational wave observables.''
  These observables are all nonlocal in time, nonzero in spacetimes with gravitational radiation, and have an observable effect that persists after the gravitational waves have passed.
  In this paper, we focus on the persistent observable known as ``curve deviation,'' and we compute the observable using the Bondi-Sachs approach to asymptotically flat spacetimes at the leading, nontrivial order in inverse Bondi radius.
  The curve deviation is related to the final separation of two observers who have an initial separation, initial relative velocity, and relative acceleration.
  The displacement gravitational wave memory effect is the part of the curve deviation that depends on the initial separation and is the entire contribution for initially comoving, inertial observers at large Bondi radius.
  The spin and center-of-mass memory effects are contained within the dependence of the curve deviation on the initial relative velocity, and the dependence of the curve deviation on relative acceleration contains observables distinct from these known memory effects.
  We find that the full curve deviation observable can be written in terms of differences in nonradiative data before and after the radiation (which we call the ``charge'' contribution), along with a nonlinear ``flux'' contribution that vanishes in the absence of gravitational radiation.
  This splitting generalizes the notion of ``ordinary'' and ``null'' memory that exists for the displacement, spin, and center-of-mass gravitational wave memory effects to the full curve deviation observable.
\end{abstract}

\maketitle

\tableofcontents

\section{Introduction}

We continue, in this paper, the investigation of ``persistent gravitational wave observables'' that was initiated in~\cite{Flanagan2019a} (henceforth, Paper I) and further developed in~\cite{Flanagan2019c} (henceforth, Paper II).
Paper I introduced persistent observables in an effort to synthesize, and generalize, an increasing number of types of gravitational wave memory effects (hereafter simply ``memory effects''\footnote{Note that there are also memory effects that occur in electromagnetism~\cite{Bieri2013a} and more general Yang-Mills theories~\cite{Pate2017}, but in this paper we will only consider memory effects due to gravitational radiation.}) that have been computed more recently (for example,~\cite{Strominger2014, Flanagan2014, Pasterski2015, Strominger2017, Nichols2018,Compere2018}) and that were not evidently related to the displacement~\cite{Zeldovich1974, Christodoulou1991} and velocity~\cite{Bondi1957, Grishchuk1989} effects that have been understood for much longer.
The persistent gravitational wave observables were constructed so as to be applicable in a number of contexts.
While the memory effect is frequently studied in asymptotically flat spacetimes (for example,~\cite{Christodoulou1991}), persistent observables are classes of permanent effects measured by idealized systems of observers, and these observables can be used in any context in which there is a well-defined notion of gravitational radiation.
These contexts include, for example, in linearized gravity on some background spacetime (which was considered, for a flat background, in Paper I), as well as in nonlinear plane-wave spacetimes (Paper II).
There were three concrete observables defined in Paper I, which involved (i) a part of the deviation vector between two accelerating curves, (ii) a holonomy observable for transport laws for linear and angular momentum, and (iii) a spinning test particle's dynamics.

Papers I and II, however, did not investigate persistent observables in much detail in the context in  which memory effects are frequently computed: in asymptotically flat spacetimes ``near'' null infinity\footnote{Because null infinity is the boundary of an asymptotically flat spacetime in which infinity has been brought to a finite point in an ``unphysical spacetime'' through the covariant conformal methods of Penrose~\cite{Penrose1962, Penrose1965}, the notion of a spacetime point being ``near'' infinity in the physical spacetime is a bit of a misnomer.
  However, the phrase is commonly used, is a convenient alternative to phrases like ``at large Bondi radius $r$'' (Bondi coordinates~\cite{Bondi1962, Sachs1962a} are defined in Sec.~\ref{sec:coordinates}), and so is one that we will use in this paper.}
at leading, nontrivial order in inverse distance (which we denote by $1/r$).
Our aim in this paper, then, is to perform this calculation so as to better understand which properties of the usual memory effects extend to more general persistent observables in this context.
Of the three persistent observables defined in Paper I, we focus on the first, which was given the name ``curve deviation.''
This observable is, of the three considered in Paper I, most closely related to the memory effect in the following sense: while the displacement memory effect is related to a change in the separation of two observers that follow geodesics and are initially comoving, the curve deviation observable also measures a change in the separation, but it does not require the observers to be geodesic or comoving.
We anticipate that to leading order in $1/r$ near null infinity, the other two persistent observables in Paper I will behave similarly to curve deviation for the following reason.
Given the scaling of curvature with $1/r$ in asymptotically flat spacetimes, we can work to linear order in the Riemann tensor; thus, the analysis in linearized gravity in Paper I (in which the three persistent observables all had qualitatively similar behaviors) also should apply to the results of this paper.
In future work, in which we plan to carry out a similar analysis to order $1/r^2$, we expect that qualitatively different features will appear in these other observables.

We now summarize the main findings of this paper.
We first compute expressions for the curve deviation at leading order in $1/r$ for a specific set of asymptotically defined observers.
Since these observers lie in the physical spacetime, we use a Bondi-Sachs expansion of the metric~\cite{Bondi1962, Sachs1962a} instead of the covariant conformal approach of Penrose~\cite{Penrose1962, Penrose1965}, although the two frameworks are equivalent, in the sense that spacetimes which are asymptotically flat in the sense of Penrose possess coordinate systems in which the metric takes the Bondi-Sachs form~\cite{Tamburino1966}.
We write our result in terms of covariant bitensors (see, for example,~\cite{Poisson2011} and Paper I) that act linearly on the initial separation between the observers, their initial relative velocity, and their initial relative acceleration (and higher derivatives, such as the initial jerk).
The dependence of the curve deviation on the initial separation contains the same information as the displacement memory effect does.
In addition, the dependence of curve deviation on the initial relative velocity is closely related to what was called the ``subleading displacement memory'' in Paper I, which contains information about the spin~\cite{Pasterski2015} and center-of-mass~\cite{Nichols2018} memory effects.
Finally, the dependence on the acceleration and its higher derivatives contains observables that are independent of these known memory effects.
All the different components of the curve deviation depend on a set of quantities that we call the \emph{temporal moments of the news tensor} (or just ``moments of the news'' for short).
Specifically, these moments of the news are retarded-time integrals of the news tensor multiplied by powers of retarded time (recall that the news tensor indicates that the spacetime is radiating gravitational radiation, since its absence indicates there is no such radiation~\cite{Bondi1962, Sachs1962a, Geroch1977}).
In particular, since the news tensor is roughly the retarded time derivative of the gravitational waveform that is measured by gravitational wave detectors, these moments are (in principle) measurable and can be computed for astrophysical sources (though we defer such calculations to future work).

We next use Einstein's equations in Bondi-Sachs form~\cite{Bondi1962, Sachs1962a} to compute the moments of the news tensor in terms of the radiative and nonradiative data that characterize an asymptotically flat spacetime, so that we can better understand what physical processes and quantities produce nontrivial values of these persistent observables.
For the displacement (and subsequently spin and center-of-mass) memory effect near null infinity in general relativity, there are several classifications of types of memory effects.
The first is the historical splitting into the \emph{linear} memory effect (named because it appears in linearized gravity~\cite{Zeldovich1974}) and the \emph{nonlinear} memory effect (which is only nonzero when nonlinear terms in Einstein's equations are taken into account~\cite{Christodoulou1991}).
A more recent classification by Bieri and Garfinkle~\cite{Bieri2013b} split the memory into \emph{ordinary} and \emph{null} parts instead.
In this taxonomy, the null memory includes contributions from all massless fields (both the radiative gravitational and matter degrees of freedom) and the ordinary part arises from changes in the data that characterize nonradiative degrees of freedom of an asymptotically flat spacetime.
Within the Bondi-Sachs approach to asymptotic flatness, it was shown, for example, in~\cite{Strominger2014, Flanagan2015} that the ordinary part of the memory corresponds to changes in the conserved quantities conjugate to the Bondi-Metzner-Sachs asymptotic symmetries~\cite{Bondi1962, Sachs1962b} and the null part corresponds to the part of the integral of the flux of these conserved charges that is nonlinear in the gravitational degrees of freedom (or in other massless fields when not in vacuum).
Thus, we adopt a third classification for the displacement, spin, and center-of-mass memory effects in asymptotically flat spacetimes in terms of their ``charge'' and ``flux'' contributions.

The second main result of this paper is that, much like with the displacement memory effect, the curve deviation observable at null infinity can be split into charge and flux contributions, though there are two important caveats that should be noted about this splitting.
The first is that the existence of ``conserved quantities'' in the charge contribution to these observables should not be taken to imply the existence of additional asymptotic symmetries: Noether's theorem implies that symmetries result in conservation laws, but the converse is not necessarily true.
Second, the flux contribution that arises for the dependence of the curve deviation on the initial relative acceleration (and higher derivatives) depends on nonradiative information that is not present in the gravitational waveform.
This is in contrast to the dependence of the curve deviation on initial separation or relative velocity, in which the flux contributions are purely radiative.

The remainder of the paper is structured as follows.
In Sec.~\ref{sec:scri}, we first review asymptotically flat spacetimes in Bondi coordinates, which we use throughout this paper.
In particular, we define the functions that occur in the asymptotic expansion of the metric (such as the mass aspect, shear, news, etc.) that are used to present our results.
We present in Sec.~\ref{sec:evolution} how these metric functions evolve with retarded time, which is a necessary result for dividing the curve deviation into its charge and flux contributions.
In Sec.~\ref{sec:persistent}, we then determine the asymptotic expansion of the curve deviation observable, and we express the observable in terms of moments of the news.
We then show in Sec.~\ref{sec:contributions} how these moments can be expressed in terms of their charge and flux contributions.
We conclude in Sec.~\ref{sec:discussion}.

We use the conventions for the signature of the metric and the Riemann tensor given in Wald~\cite{Wald1984}.
Lowercase Latin letters from the beginning of the alphabet will be used for abstract spacetime indices.
Our conventions for quantities in Bondi coordinates are based upon those in~\cite{Flanagan2015}, although we make a handful of minor changes that are given in Sec.~\ref{sec:coordinates}.
In particular, we use lowercase Latin letters from the middle of the alphabet ($i$, $j$, etc.) for the angular coordinate indices, instead of the (conventional) uppercase Latin letters from the beginning of the alphabet, and use $\ms D_i$ to denote the covariant derivative on the unit two-sphere instead of $D_A$.
Our conventions for bitensors follow those of the review article~\cite{Poisson2011}, and we use the convention that indices at a point $x$ are denoted $a, b, \ldots$, while those at $x'$ are denoted $a', b', \ldots$, etc.
Following Paper I, we use Greek indices to indicate components along a parallel-transported tetrad.
Finally, also following Paper I, we take powers of order symbols, writing (for example) $O(a, b)^2 \equiv O(a^2, ab, b^2)$, for brevity.

\section{Review of asymptotically flat spacetimes} \label{sec:scri}

\subsection{Bondi coordinates and metric} \label{sec:coordinates}

We use Bondi coordinates $u$, $r$, $\theta^i$, where $u$ is a retarded time coordinate, $r$ is a radial variable, and $\theta^i$ are angular coordinates on the unit two-sphere.
Because we will focus on asymptotically flat solutions, we will write the metric in the following form, which differs slightly from the standard parametrization of the Bondi-Sachs metric:
\begin{equation} \label{eqn:metric}
  \begin{split}
    \ud s^2 &= -\left(1 - \frac{2V}{r}\right) e^{2 \beta/r} \ud u^2 - 2 e^{2 \beta/r} \ud u \ud r \\
    &\hspace{1em}+ r^2 \mc H_{ij} \left(\ud \theta^i - \frac{\mc U^i}{r^2} \ud u\right) \left(\ud \theta^j - \frac{\mc U^j}{r^2} \ud u\right).
  \end{split}
\end{equation}
Note that the metric components satisfy $g_{rr} = 0$ and $g_{ri} = 0$, which fix three of the four gauge degrees of freedom of the metric.
The final gauge degree of freedom is fixed by the condition that
\begin{equation} \label{eqn:gauge}
  \partial_r \det \mc H = 0.
\end{equation}
These choices set up a Bondi coordinate system~\cite{Bondi1962, Sachs1962a}.

The Bondi metric in Eq.~\eqref{eqn:metric} is frequently (though need not only be) used to describe asymptotically flat solutions.
To have the metric~\eqref{eqn:metric} correspond to such a solution, it is necessary to impose boundary conditions on the metric functions $V$, $\beta$, $\mc U^i$, and $\mc H_{ij}$ so that the metric approaches a Minkowski form as $r \to \infty$.
Given our parametrization of the metric in Eq.~\eqref{eqn:metric}, the metric functions $V$, $\beta$, $\mc U^i$, and $\mc H_{ij}$ must all be $O(1)$.
We will also require that these functions admit a power series expansion in $1/r$ (that is, without terms involving $\log r$).
Finally, we will write
\begin{equation}
  \mc H_{ij} = h_{ij} + O(1/r),
\end{equation}
where $h_{ij}$ is the usual metric on the unit two-sphere.
We use $h_{ij}$ to raise and lower the two-sphere indices (that is, $i$, $j$, etc.).
For later convenience, we also define $\ms D_i$ to be the connection on the two-sphere that is compatible with $h_{ij}$, and we define $\epsilon_{ij}$ to be the volume form on the two-sphere.

With this setup, one can perform a fairly standard calculation (see, for example,~\cite{Flanagan2015, Madler2016}) to show that the Einstein equations, in vacuum, give rise to a set of partial differential equations in $r$ (the \emph{hypersurface equations}) for $V$, $\beta$, and $\mc U^i$, which can be solved hierarchically in terms of $\mc H_{ij}$.
We review more of the details of this calculation in Appendix~\ref{app:hyper}.
The solutions of these equations imply that
\begin{subequations}
  \begin{align}
    \beta &= \tilde{\beta}/r, \label{eqn:beta_exp} \\
    V &= m + \mc M/r, \label{eqn:V_exp} \\
    \mc U^i &= U^i + V^i/r + \Upsilon^i/r^2, \label{eqn:U_exp}
  \end{align}
\end{subequations}
where $U^i$ will be determined by the $O(1/r)$ coefficient in the expansion of $\mc H_{ij}$, where $m$ and $V^i$ are functions of integration (that is, functions of $u$ and $\theta^i$) that arise from solving these radial partial differential equations, and where $\tilde{\beta}$, $\mc M$, and $\Upsilon^i$ are some $O(1)$ functions that admit a Taylor series expansion in $1/r$ and are determined by the radial differential equations.
The exact forms of $\tilde{\beta}$, $\mc M$, and $\Upsilon^i$ are not relevant here, but can be determined from Eqs.~\eqref{eqn:hyper_beta}, \eqref{eqn:hyper_V}, and~\eqref{eqn:hyper_U}, respectively.

Finally, we consider the expansion of $\mc H_{ij}$.
First, the gauge condition~\eqref{eqn:gauge} implies that \{see, for example, Eq.~(12) of~\cite{Barkett2019}\}
\begin{subequations}
  \begin{align}
    \mc H_{ij} &= \sqrt{1 + \frac{\mc C_{kl} \mc C^{kl}}{2r^2}} h_{ij} + \frac{1}{r} \mc C_{ij}, \\
    (\mc H^{-1})^{ij} &= \sqrt{1 + \frac{\mc C_{kl} \mc C^{kl}}{2r^2}} h^{ij} - \frac{1}{r} \mc C^{ij},
  \end{align}
\end{subequations}
where $\mc C_{ij}$ is trace-free with respect to $h_{ij}$, and where we have used the notation $(\mc H^{-1})^{ij}$ to indicate that this tensor is the inverse of $\mc H_{ij}$, and not $\mc H_{ij}$ with its indices raised by $h^{ij}$.
This result can be verified by using Jacobi's determinant formula and the relationship that $\mc C_{ik} \mc C^{kj} = \frac{1}{2} \mc C_{kl} \mc C^{kl} \delta_i{}^j$ for any symmetric, trace-free tensor $\mc C_{ij}$ on the two-sphere.

We therefore find that the expansion of $\mc H_{ij}$ is entirely determined by the expansion of $\mc C_{ij}$.
The requirement that no $\log r$ terms occur when solving the hypersurface equation for $\mc U^i$ imposes that the coefficient of the $1/r$ term in the power-series expansion of $\mc C_{ij}$ vanishes, so that
\begin{equation} \label{eqn:C_exp}
  \mc C_{ij} = C_{ij} + \frac{1}{r^2} \mc E_{ij}.
\end{equation}
The function $C_{ij}$ is constant in $r$ (that is, depends on just $u$ and $\theta^i$) and $\mc E_{ij}$ is $O(1)$ and has an expansion of the form
\begin{equation} \label{eqn:calEij}
  \mc E_{ij} = \sum_{n = 0}^\infty \frac{1}{r^n} \ord{n}{\mc E}_{ij}.
\end{equation}

In terms of the expansion of $\mc H_{ij}$, the quantity $U^i$ in the expansion of $\mc U^i$ can be shown from the hypersurface equations to be
\begin{equation}
  U^i = -\frac{1}{2} \ms D_j C^{ij}.
\end{equation}
We also write $V^i$ as
\begin{equation} \label{eqn:U_1}
  V^i \equiv -\frac{2}{3} N^i + \frac{1}{16} \ms D^i (C_{jk} C^{jk}) + \frac{1}{2} C^{ij} \ms D^k C_{jk}.
\end{equation}
This is purely convention: $V^i$ and $N^i$ are entirely equivalent.

In summary, the metric above can be written entirely in terms of the quantities $m$, $N^i$, $C_{ij}$, and $\mc E_{ij}$.
The first three of these quantities have names: $m$ is known as the mass aspect, $N^i$ as the angular momentum aspect, and $C_{ij}$ as the shear.
These three quantities are purely functions of $u$ and $\theta^i$.
The quantity $\mc E_{ij}$ is not named, and it is a function of all four Bondi coordinates, as shown in Eq.~\eqref{eqn:calEij}.

\subsection{Geodesic equation} \label{sec:geodesic}

For computing the curve deviation observable, which involves timelike, accelerating worldlines near null infinity, it will be helpful to first review the behavior of geodesics in asymptotically flat spacetimes at large Bondi radius.
Consider a timelike worldline $\gamma$.
We will write the four-velocity $\dot{\gamma}^a$ of this worldline as
\begin{equation}
  \dot{\gamma}^a = \chi (\partial_u)^a + \frac{1}{r} v^i (\partial_i)^a + \dot{r} (\partial_r)^a,
\end{equation}
where the normalization $\dot{\gamma}^a \dot{\gamma}_a = -1$ of the four-velocity implies that
\begin{equation} \label{eqn:dot_r}
  \begin{split}
    \dot{r} = \frac{1}{2\chi} e^{-2 \beta/r} \bigg[1 &+ \mc H_{ij} \left(v^i - \frac{\chi}{r} \mc U^i\right) \left(v^j - \frac{\chi}{r} \mc U^j\right) \\
    &- \left(1 - \frac{2V}{r}\right) e^{2 \beta/r} \chi^2\bigg].
  \end{split}
\end{equation}
We assume that the quantities $\chi$ and $v^i$ are $O(1)$.
In particular, we consider $v^i$ as $O(1)$ (as opposed to $\dot{\theta}^i = v^i/r$) because physical distances between points at different values of $\theta^i$ go as $r \Delta \theta^i$; thus, $\dot{\theta}^i$ being $O(1)$ implies an observer moving at infinite velocity at infinity.

The geodesic equation, in terms of $\chi$ and $v^i$, becomes
\begin{equation} \label{eqn:dot_chi_v}
  \dot{\chi} = O(1/r), \qquad \dot{v}^i = O(1/r).
\end{equation}
This calculation requires the Christoffel symbols for the metric~\eqref{eqn:metric}, or rather the orders of the Christoffel symbols in an expansion in $1/r$.
We give these in Appendix~\ref{app:Christoffel}.
To leading order, it is reasonable to consider both $\chi$ and $v^i$ to be constant.
In this paper, we will typically consider the case where
\begin{equation} \label{eqn:obs_worldline}
  \chi = 1 + O(1/r), \qquad v^i = O(1/r),
\end{equation}
in which case we have that
\begin{equation}
  \dot{r} = O(1/r).
\end{equation}
As such, these curves (to leading order) are curves of constant $r$.

A similar calculation that requires the Christoffel symbols is that of the parallel transport of the following quantities:
\begin{equation}
  (\hat{\theta}_i)^a \equiv \frac{1}{r} (\partial_i)^a, \qquad (\hat{\theta}^i)_a \equiv r (\ud \theta^i)^a.
\end{equation}
These are versions of the vector field $(\partial_i)^a$ and one-form $(\ud \theta^i)_a$, respectively, that have finite magnitude at infinity.
These quantities are parallel-transported to leading order in $1/r$:
\begin{equation} \label{eqn:dyad_prop}
  \dot{\gamma}^b \nabla_b (\hat{\theta}_i)^a = O(1/r), \qquad \dot{\gamma}^b \nabla_b (\hat{\theta}^i)_a = O(1/r).
\end{equation}
The left-hand sides of these equations are tensorial quantities; the meaning of the $O(1/r)$ on the right-hand sides is that these tensors have $O(1/r)$ components on the tetrad given by $\{\partial_u, \partial_r, \hat{\theta}_1, \hat{\theta}_2\}$ and $\{\ud u, \ud r, \hat{\theta}^1, \hat{\theta}^2\}$.

\subsection{Riemann tensor}

For a vacuum spacetime, different components of the Riemann (or equivalently Weyl) tensor can be written in terms of an expansion in $1/r$ and in terms of the Bondi metric functions in Sec.~\ref{sec:coordinates}.
To give all the relevant components that we use in this paper, we first need to introduce the news tensor $N_{ij} = \partial_u C_{ij}$, which will be described in greater detail in Sec.~\ref{sec:evolution}.
The first set of components of the Riemann tensor is given by
\begin{equation} \label{eqn:Riemann}
  R^i{}_{uju} = -\frac{1}{2r} \partial_u N^i{}_j + O(1/r^2),
\end{equation}
which is the leading order contribution to the Riemann tensor:
\begin{equation}
  R^a{}_{bcd} = (\hat{\theta}_i)^a (\partial_u)_b (\hat{\theta}^j)_c (\partial_u)_d R^i{}_{uju} + O(1/r^2)
\end{equation}
[where the $O(1/r^2)$ has the same meaning as in Eq.~\eqref{eqn:dyad_prop} above].
The fact that the leading-order piece of the Riemann tensor is entirely given by the news is well known (for example,~\cite{Bondi1962, Sachs1962a}), but it is a key result for computing the curve deviation observable in this paper.

Another feature of the Riemann tensor in Bondi coordinates is that, in linearized gravity, the quantities $m$, $N_i$, and $\mc E_{ij}$ each show up in certain distinct components of this tensor.
The relevant components are $R_{urur}$, $R_{urri}$, and $R_{rirj}$:
\begin{subequations} \label{eqn:Riemann_nonradiative}
  \begin{align}
    R_{urur} &\simeq -\partial_r^2 \left(\frac{V}{r}\right) = -\frac{2m}{r^3} + O(1/r^4), \\
    R_{urri} &\simeq -\frac{r}{2} \partial_r \left(\frac{\partial_r \mc U_i}{r}\right) = \frac{N_i}{r^3} + O(1/r^4), \\
    R_{rirj} &\simeq -\frac{1}{2} r \partial_r^2 \left(\frac{\mc E_{ij}}{r^2}\right) = -\frac{1}{r^3} \sum_{n = 0}^\infty \frac{(2 + n) (3 + n)}{2} \frac{\ord{n}{\mc E}_{ij}}{r^n}. \label{eqn:Riemann_E}
  \end{align}
\end{subequations}
Here ``$\simeq$'' indicates an equality that holds in the linearized approximation, where we neglect terms that are quadratic in $m$, $N_i$, $C_{ij}$, $\mc E_{ij}$ and their derivatives.
These relationships between the metric functions $m$, $N_i$, and $\mc E_{ij}$ and the curvature provide some additional insight into the physical properties of these metric functions.
For example, for some linearized source, one can compute the values of the Riemann tensor far from this source, and then use Eq.~\eqref{eqn:Riemann_nonradiative} in order to relate integrals of the stress-energy tensor of the source to $m$, $N_i$, and $\mc E_{ij}$.
This provides one motivation for the names ``mass aspect'' and ``angular momentum aspect'' for $m$ and $N_i$, respectively.
Another method, using coordinate transformations, and allowing for an analysis at the nonlinear level, was performed in~\cite{Blanchet2020}.

\subsection{Conservation and evolution equations} \label{sec:evolution}

The above analysis of the metric implies that the metric on a hypersurface of constant $u$ can be determined entirely in terms of the functions of integration, $m$ and $N_i$, as well as the quantities $C_{ij}$ and $\mc E_{ij}$.
However, we have not specified how these quantities can be computed at future values of $u$ from their values at some known value of $u$.
There are other components of Einstein's equations, which we have not yet used, that describe how $m$, $N_i$, $C_{ij}$ and $\mc E_{ij}$ evolve.

The trace-free part (with respect to $\mc H_{ij}$) of the $ij$ components of the Einstein equations determines how $C_{ij}$ and $\mc E_{ij}$ evolve, and these components are frequently referred to as the ``evolution equations'' in the Bondi-Sachs framework~\cite{Madler2016}.
These equations can be written such that $\partial_u \mc H_{ij}$ obeys a differential equation in $r$ with a right-hand side that depends explicitly (and implicitly) on $\mc H_{ij}$, as well as on $m$ and $N_i$ [see Eq.~\eqref{eqn:hyper_dHdu}].
The solution to this differential equation [see Eq.~\eqref{eqn:dot_C}] introduces a new set of functions of integration, denoted $N_{ij}$, which are related to the tensor $C_{ij}$ in the expansion of $\mc H_{ij}$ by
\begin{equation} \label{eqn:news}
  N_{ij} \equiv \partial_u C_{ij}.
\end{equation}
This quantity is known as the (Bondi) \emph{news tensor} (or just ``news'') and is free data in this problem, in the sense that $N_{ij}$ is an unconstrained function of $u$ and $\theta^i$.
It has the property that it vanishes in stationary regions of a spacetime (see, for example,~\cite{Bondi1962}); however, a region where the news vanishes is not necessarily stationary.
Regions with vanishing news provide a notion of a ``nonradiative'' region of the spacetime; for example, in linearized gravity in such regions, there would be no gravitational waves.

The higher order in $1/r$ parts of the Einstein equations involving $\partial_u \mc H_{ij}$ fix $\partial_u \mc E_{ij}$ in terms of $m$, $N_i$, $C_{ij}$, $\mc E_{ij}$ and their derivatives.
The evolution equation for $\mc E_{ij}$ proves to be much more complex in general,
but the evolution equation for the leading order piece, $\ord{0}{\mc E}_{ij}$ (given in, for example,~\cite{Nichols2018}) takes the form
\begin{equation} \label{eqn:dot_E0}
  \begin{split}
    \partial_u \ord{0}{\mc E}_{ij} &= \frac{1}{4} N_{kl} C^{kl} C_{ij} + \frac{1}{3} \STF \ms D_i N_j \\
    &\hspace{1em}+ \frac{1}{4} C_j{}^k \ms D_{[i} \ms D^l C_{k]l} + \frac{1}{2} m C_{ij}.
  \end{split}
\end{equation}
Here, we have introduced the notation $\STF$, which means to take the symmetric trace-free part of the free indices in the expression (where ``trace-free'' means with respect to the metric $h_{ij}$).
The subleading order piece, $\ord{1}{\mc E}_{ij}$, obeys a similar evolution equation that is given by~\cite{vanderBurg1966, Godazgar2018a}
\begin{equation} \label{eqn:dot_E1}
  \begin{split}
    \partial_u \ord{1}{\mc E}_{ij} &= -\frac{1}{4} (\ms D^2 + 2) \ord{0}{\mc E}_{ij} \\
    &\hspace{1em}- \STF \ms D^k \bigg\{\bigg[\frac{3}{32} \ms D_i (C_{lm} C^{lm}) - \frac{1}{4} C_{il} \ms D_m C^{lm} \\
    &\hspace{6.9em}+ \frac{1}{3} N_i\bigg] C_{jk} - \frac{5}{32} C_{lm} C^{lm} \ms D_i C_{jk}\bigg\}.
  \end{split}
\end{equation}
The evolution equations for $\ord{n + 1}{\mc E}_{ij}$, for $n \geq 1$, are similar, in the sense that their linear piece involves only an operator acting on $\ord{n}{\mc E}_{ij}$:
\begin{equation} \label{eqn:dot_En}
  \partial_u \ord{n + 1}{\mc E}_{ij} \simeq \mc D_n \ord{n}{\mc E}_{ij},
\end{equation}
where
\begin{equation} \label{eqn:D_n_def}
  \mc D_n \equiv -\frac{n + 2}{2(n + 1) (n + 4)} \left(\ms D^2 + n^2 + 5n + 2\right),
\end{equation}
and [as in Eq.~\eqref{eqn:Riemann_nonradiative}] we use ``$\simeq$'' to indicate that this expression only contains the \emph{linear} terms.
We provide a schematic form of the full evolution equation for $\ord{n + 1}{\mc E}_{ij}$ below, in Eq.~\eqref{eqn:general_evol}, which can be determined from Eq.~\eqref{eqn:dot_C}.
Equation~\eqref{eqn:dot_En} can be derived from the linearization of Eq.~\eqref{eqn:dot_C}, as outlined at the end of Appendix~\ref{app:hyper}.

The evolution of $m$ and $N_i$ is determined by the $uu$ and $ui$ components of Einstein's equations, which are referred to as the ``conservation equations'' (see, for example,~\cite{Madler2016}).
They are given by
\begin{subequations}
  \begin{align}
    \partial_u m &= \frac{1}{4} \left(\ms D_i \ms D_j N^{ij} - \frac{1}{2} N_{ij} N^{ij}\right), \label{eqn:dot_m} \\
    \partial_u N_i &= \ms D_i m - \frac{1}{4} \epsilon_{ji} \ms D^j (\epsilon^{kl} \ms D_k \ms D^m C_{lm}) \nonumber \\
    &\hspace{1em}+ \frac{1}{4} \left(N^{jk} \ms D_j C_{ki} + 3 C_{ij}\ms D_k N^{jk}\right). \label{eqn:dot_N}
  \end{align}
\end{subequations}

In summary, therefore, the situation is even simpler than that presented in Sec.~\ref{sec:coordinates}: the metric is entirely determined by the initial values of $m$, $N_i$, and $\mc E_{ij}$ at some value of $u$, along with the value of $C_{ij}$ at each value of $u$.
As such, we split these quantities into radiative and nonradiative degrees of freedom: $m$, $N_i$, and $\mc E_{ij}$ are nonradiative, whereas $C_{ij}$ is radiative.\footnote{Another way of splitting these quantities is to say that only $N_{ij}$ contains radiative degrees of freedom and to append Eq.~\eqref{eqn:news} to our list of evolution equations.
  In this splitting, $C_{ij}$ is nonradiative, and requires its initial value in order to determine its value at all later times.
  This splitting is sensible, since it is only $N_{ij}$ that characterizes the presence of radiation.
  Our choice of calling $C_{ij}$ radiative is motivated by the inclusion of the shear in the radiative phase space of~\cite{Ashtekar1981, Ashtekar2014}, as well as the fact that it is $C_{ij}$ which gravitational wave interferometers measure, and not $N_{ij}$.}

\section{Curve deviation observable} \label{sec:persistent}

\subsection{Definition and properties}

We review, in this section, the curve deviation persistent observable that was given in Paper I.
This observable is defined by the following procedure.
Suppose there are two observers who follow the two worldlines $\gamma$ and $\bar{\gamma}$, respectively.
For simplicity, we assume that $\gamma$ is geodesic, but that $\bar{\gamma}$ has a nonzero acceleration $\ddot{\bar \gamma}^{\bar a}$ (note that we place an overline on the indices of points along $\bar{\gamma}$).
We parametrize each of these worldlines by proper time $\tau$, and we choose to set the proper times of these two worldlines to be equal to some common value $\tau_0$ at some initial point $x$ along $\gamma$ and $\bar x$ along $\bar{\gamma}$.
At $x$, we consider a separation vector $\xi^a$ and relative velocity vector $\dot{\xi}^a$; the former is defined by the exponential map (for details, see~\cite{Vines2014a}) and the latter is defined by taking a covariant derivative of $\xi^a$ with respect to $\tau$.

Next, the separation $\xi^a$ between the worldlines obeys the following differential equation~\cite{Flanagan2019a}:
\begin{equation} \label{eqn:non_geodesic}
  \ddot{\xi}^a = -R^a{}_{cbd} \dot{\gamma}^c \dot{\gamma}^d \xi^b + a^a,
\end{equation}
where $\ddot{\xi}^a$ is the derivative of $\dot{\xi}^a$ with respect to $\tau$, and
\begin{equation}
  a^a \equiv g^a{}_{\bar a} \ddot{\bar \gamma}^{\bar a}
\end{equation}
is a sort of relative acceleration vector between the two worldlines.
In this expression, $g^a{}_{\bar a}$ is the parallel propagator between $\gamma(\tau)$ and $\bar{\gamma} (\tau)$ (for arbitrary $\tau$), parallel-transporting vectors along the unique geodesic\footnote{This geodesic always exists for $\gamma$ and $\bar{\gamma}$ sufficiently closely-separated; such an assumption is built into this discussion, as $\xi^a$ is only defined in such a case.} between $\gamma(\tau)$ and $\bar{\gamma} (\tau)$.
Note that, due to the presence of $a^a$ on the right-hand side of Eq.~\eqref{eqn:non_geodesic}, the final separation between the observers at some later proper time $\tau_1$ is nonvanishing, even in the absence of curvature.
In the absence of curvature, the final separation is the same as the initial separation in the case where the observers have zero acceleration and are initially comoving; a nonzero initial velocity or relative acceleration will lead to a different final separation.
Because we are primarily interested in the effects of curvature on the final separation, we subtract the solution to the following differential equation:
\begin{equation} \label{eqn:flat_deviation}
  \ddot{\xi}_{\rm flat}^a = a^a.
\end{equation}
The \emph{curve deviation observable} then was defined in~\cite{Flanagan2019a} to be
\begin{equation} \label{eqn:Delta_xi}
  \Delta \xi^{a'} \equiv \xi^{a'} - \xi_{\rm flat}^{a'}.
\end{equation}
All of these quantities are computed at $\gamma(\tau_1) \equiv x'$, where $\tau_1 > \tau_0$ (recall that we are parametrizing both of these worldlines by the \emph{same} affine parameter).

To determine the curve deviation observable, therefore, in addition to determining the final separation from Eq.~\eqref{eqn:non_geodesic}, we also need to compute the solution to Eq.~\eqref{eqn:flat_deviation}.
The solution to Eq.~\eqref{eqn:flat_deviation} can be straightforwardly verified to be
\begin{equation}
  \begin{split}
    \xi^{a'}_{\rm flat} &\equiv \pb{\gamma} g^{a'}{}_a [\xi^a + (\tau_1 - \tau_0) \dot{\xi}^a] \\
    &\hspace{1em}+ \int_{\tau_0}^{\tau_1} \ud \tau_2\; (\tau_1 - \tau_2) \pb{\gamma} g^{a'}{}_{a''} a^{a''},
  \end{split}
\end{equation}
where $x'' = \gamma(\tau_2)$ and $\pb{\gamma} g^{a'}{}_a$ is the parallel propagator between two points $x$ and $x'$ on a curve $\gamma$.
In the second line, we have used the Cauchy rule for repeated integration to write what naturally is a double integral in terms of a single integral.

The curve deviation observable can then be parametrized in terms of its dependence on $\xi^a$, $\dot{\xi}^a$, and
\begin{equation}
  \ord{n} a^a \equiv \left.\frac{\uD^n a^a}{\ud \tau^n}\right|_{\tau = \tau_0},
\end{equation}
as follows~\cite{Flanagan2019a}:
\begin{equation} \label{eqn:Delta_xi_exp}
  \begin{split}
    \Delta \xi^{a'} &\equiv \pb{\gamma} \Delta K^{a'}{}_a \xi^a + \pb{\gamma} \Delta H^{a'}{}_a \dot{\xi}^a + \sum_{n = 0}^\infty \pb{\gamma} \Delta \ord{n} \alpha^{a'}{}_a \ord{n} a^a \\
    &\hspace{1em}+ O(\bs \xi, \dot{\bs \xi}, \bs a)^2.
  \end{split}
\end{equation}
That is, the quantities $\pb{\gamma} \Delta K^{a'}{}_a$, $\pb{\gamma} \Delta H^{a'}{}_a$, and $\pb{\gamma} \Delta \ord{n} \alpha^{a'}{}_a$ describe the dependence of the curve deviation observable on initial values of the separation, relative velocity, and the $n$th time derivative of $a^a$, respectively.
All three of these quantities vanish in the absence of curvature due to the subtraction of $\xi_{\rm flat}^{a'}$ in Eq.~\eqref{eqn:Delta_xi}.
Since Eq.~\eqref{eqn:non_geodesic} is the geodesic deviation equation, but with an extra ``source'' term on the right-hand side, all three of these quantities can be written in terms of the solutions to the geodesic deviation equation, which are known as \emph{Jacobi propagators}.
For more details, see Paper I; we will not require the formalism of Jacobi propagators in this paper.

Finally, note that (as defined) $a^a$ is difficult to measure, as its measurement by the observer along $\gamma$ would require the parallel transport of $\ddot{\bar \gamma}^{\bar a''}$ between the two geodesics at all values of $\tau_2$ between $\tau_0$ and $\tau_1$.
In Appendix~\ref{app:rel_acc}, we show that
\begin{equation} \label{eqn:rel_acc}
  \ord{n} a^a = g^a{}_{\bar a} \left.\frac{\uD^n \ddot{\bar \gamma}^{\bar a}}{\ud \tau^n}\right|_{\tau = \tau_0} + O(\bs \xi, \dot{\bs \xi}, \ddot{\bar{\bs \gamma}})^2.
\end{equation}
This shows that $\Delta \ord{n} \alpha^{a'}{}_a$ describes the dependence of the curve deviation on the $n$th time derivative of $\ddot{\bar \gamma}^{\bar a}$ at $\tau = \tau_0$, providing a clearer physical interpretation of these terms.

\subsection{Asymptotic expansions and moments of the news} \label{sec:moments}

We now consider asymptotic expansions of the bitensors that characterize the curve deviation observable, $\pb{\gamma} \Delta K^{a'}{}_a$, $\pb{\gamma} \Delta H^{a'}{}_a$, and $\pb{\gamma} \Delta \ord{n} \alpha^{a'}{}_a$.
To do so, we use the results from Paper I, which gives these observables in terms of an expansion in powers of the Riemann tensor along a parallel-transported basis:
\begin{widetext}
\begin{subequations} \label{eqn:obs_Riemann}
  \begin{align}
    \pb{\gamma} \Delta K^\alpha{}_\beta (\tau_1, \tau_0) &= -\int_{\tau_0}^{\tau_1} \ud \tau_2 \int_{\tau_0}^{\tau_2} \ud \tau_3 R^\alpha{}_{\gamma \beta \delta} (\tau_3) \dot{\gamma}^\gamma \dot{\gamma}^\delta + O(\bs R^2), \label{eqn:DeltaK_Riemann} \\
    (\tau_1 - \tau_0) \pb{\gamma} \Delta H^\alpha{}_\beta (\tau_1, \tau_0) &= -\int_{\tau_0}^{\tau_1} \ud \tau_2 \int_{\tau_0}^{\tau_2} \ud \tau_3 (\tau_3 - \tau_0) R^\alpha{}_{\gamma \beta \delta} (\tau_3) \dot{\gamma}^\gamma \dot{\gamma}^\delta + O(\bs R^2), \label{eqn:DeltaH_Riemann} \\
    \pb{\gamma} \Delta \ord{n} \alpha^\alpha{}_\beta (\tau_1, \tau_0) &= -\frac{1}{n!} \int_{\tau_0}^{\tau_1} \ud \tau_2 (\tau_2 - \tau_0)^n \int_{\tau_2}^{\tau_1} \ud \tau_3 \int_{\tau_2}^{\tau_3} \ud \tau_4 (\tau_4 - \tau_2) R^\alpha{}_{\gamma \beta \delta} (\tau_4) \dot{\gamma}^\gamma \dot{\gamma}^\delta + O(\bs R^2). \label{eqn:Deltaalpha_Riemann}
  \end{align}
\end{subequations}
\end{widetext}
In this expression, Greek letter indices indicate components on a parallel-transported basis (note that $\dot{\gamma}^\alpha$ is not a function of $\tau$, since $\gamma$ is assumed to be a geodesic).

We assume henceforth that the news $N_{ij}$ vanishes when $u \leq u_0$ and $u \geq u_1$, where $u_0$ and $u_1$ is our notation for the values of $u$ at $x$ and $x'$, respectively.
By Eqs.~\eqref{eqn:dot_m}, \eqref{eqn:dot_N}, and~\eqref{eqn:dot_En}, it follows that the metric functions $m$, $N_i$, and $\ord{n}{\mc E}_{ij}$ all become polynomial in $u$ outside of $[u_0, u_1]$.\footnote{Specifically, $m$ is a constant, $N_i$ is a linear function of $u$, and $\ord{n}{\mc E}_{ij}$ is a polynomial of degree $n + 2$.}
While this assumption may not be representative of the class of asymptotically flat spacetimes, where the news $N_{ij}$ is typically assumed to fall off as $1/|u|^{1 + \epsilon}$, for some $\epsilon > 0$ (see, for example,~\cite{Prabhu2019}), we anticipate that removing this assumption will not qualitatively change most of our results.
The clearest difference is that having nonvanishing news outside the interval $[u_0, u_1]$ would primarily introduce terms proportional to the value of the news at $u_0$ and $u_1$ into our expressions below.

To evaluate Eq.~\eqref{eqn:obs_Riemann} in asymptotically flat spacetimes, we use Eq.~\eqref{eqn:Riemann} for the leading-order Riemann tensor, and then we make the following assumptions.
First, we suppose that our observer has a four-velocity $\dot{\gamma}^a = (\partial_u)^a + O(1/r)$, or, in other words, that Eq.~\eqref{eqn:obs_worldline} holds.
Second, for three of the members of our parallel-transported basis of vectors, we use $\dot{\gamma}^a$ and $(\hat{\theta}_i)^a$, and for our parallel-transported basis of one-forms, we use $\dot{\gamma}_a$ and $(\hat{\theta}^i)_a$.
The facts that $\dot\gamma^a$ is tangent to a geodesic and that $(\hat{\theta}_i)^a$ and $(\hat{\theta}^i)_a$ are parallel transported along this geodesic to leading order in $1/r$ were shown in Sec.~\ref{sec:geodesic}.
For simplicity, we do not consider the components of our observables along the fourth member of these bases (which would correspond to the radial direction).
Moreover, note that the components of our observables along $\dot{\gamma}^a$ vanish by the symmetries of the Riemann tensor and Eq.~\eqref{eqn:obs_Riemann}.

With these assumptions, we now compute the components of $\pb{\gamma} \Delta K^\alpha{}_\beta (\tau_1, \tau_0)$, $\pb{\gamma} \Delta H^\alpha{}_\beta (\tau_1, \tau_0)$, and $\pb{\gamma} \ord{n}{\Delta \alpha}^\alpha{}_\beta (\tau_1, \tau_0)$ along $(\hat{\theta}_i)^a$ and $(\hat{\theta}^i)_a$, which we denote by $\Delta K^i{}_j (u_1, u_0)$, $\Delta H^i{}_j (u_1, u_0)$, and $\ord{n}{\Delta \alpha}^i{}_j (u_1, u_0)$, respectively.
First, we find that Eq.~\eqref{eqn:DeltaK_Riemann} becomes
\begin{equation}
  \begin{split}
    \Delta K^i{}_j (u_1, u_0) &= \frac{1}{2r} \int_{u_0}^{u_1} \ud u_2 \int_{u_0}^{u_2} \ud u_3 \dot{N}^i{}_j (u_3) + O(1/r^2) \\
    &= \frac{1}{2r} \int_{u_0}^{u_1} \ud u_2 N^i{}_j (u_2) + O(1/r^2),
  \end{split}
\end{equation}
where we have used the fact that the news tensor vanishes at $u_0$ and $u_1$.
To help make the following discussion more systematic, we will define the $n$th moment of the news by
\begin{equation} \label{eqn:moment}
  \ord{n}{\mc N}{}^i{}_j (u_1, u_0) \equiv \frac{1}{n!} \int_{u_0}^{u_1} \ud u_2 (u_2 - u_0)^n N^i{}_j (u_2).
\end{equation}
Note that, since we assume that $N_{ij}$ vanishes outside of $[u_0, u_1]$, $\ord{n}{\mc N}{}^i{}_j (u, u_0)$ is constant in $u$ for $u > u_1$.
Using Eq.~\eqref{eqn:moment}, we can write the expression for $ \Delta K^i{}_j (u_1, u_0)$ as
\begin{equation} \label{eqn:DeltaK_moment}
  \Delta K^i{}_j (u_1, u_0) = \frac{1}{2r} \ord{0}{\mc N}{}^i{}_j (u_1, u_0) + O(1/r^2).
\end{equation}
Note that the zeroth moment of the news, $\ord{0}{\mc N}{}^i{}_j (u_1, u_0)$, is just the difference of the components of the shear $C{}^i{}_j$ between the times $u_0$ and $u_1$.

After several integrations by parts, which are described in Appendix~\ref{app:ibp}, analogous versions of Eq.~\eqref{eqn:DeltaK_moment} can be derived, so as to write $\Delta H^i{}_j (u_1, u_0)$ and $\Delta \ord{n} \alpha^i{}_j (u_1, u_0)$ in terms of moments of the news.
They are given by
\begin{equation} \label{eqn:DeltaH_moment}
  \begin{split}
    (u_1 - u_0) \Delta H^i{}_j (u_1, u_0) &= \frac{1}{r} \bigg[\ord{1}{\mc N}{}^i{}_j (u_1, u_0) \\
    &\hspace{2.5em}- \frac{1}{2} (u_1 - u_0) \ord{0}{\mc N}{}^i{}_j (u_1, u_0)\bigg] \\
    &\hspace{1em}+ O(1/r^2)
  \end{split}
\end{equation}
and
\begin{equation} \label{eqn:Deltaalpha_moment}
  \begin{split}
    \Delta \ord{n} \alpha^i{}_j (u_1, u_0) &= \frac{1}{2r} \bigg[(n + 3) \ord{n + 2}{\mc N}{}^i{}_j (u_1, u_0) \\
    &\hspace{3em}- (u_1 - u_0) \ord{n + 1}{\mc N}{}^i{}_j (u_1, u_0)\bigg] \\
    &\hspace{1em}+ O(1/r^2).
  \end{split}
\end{equation}
These three equations provide the relationships between the various pieces of the curve deviation observables and moments of the news, for the particular choice of asymptotic observers described in this section.
The remainder of this paper derives expressions for the moments of the news, because the curve deviation observables can be determined from the moments straightforwardly.

\section{``Charge'' and ``flux'' contributions to the moments of the news} \label{sec:contributions}

In this section, we describe a splitting of the moments of the news into parts that we call ``charge'' and ``flux'' contributions, because the parts have some similarities to the charge and flux parts of the displacement, spin, and center-of-mass memory effects (though some notable differences also arise).
For the first few lowest moments of the news, we compute these charge and flux terms by using the evolution equations given in Sec.~\ref{sec:evolution} for $m$, $N_i$, and $\ord{n}{\mc E}_{ij}$ (for $n=0$ and 1).
For $n > 1$, we instead use the evolution equation in Appendix~\ref{app:hyper} to infer the form of these two types of contributions to the curve deviation observable, but we do not give explicit expressions for these contributions.
This section is organized such that the results are given in the first subsection, the derivation of the results are in the second subsection, and some properties of the multipolar expansion of the temporal moments of the news are given in the final subsection.

\subsection{Summary of the moments}

The zeroth moment of the news---or equivalently the change in the shear, the displacement memory effect, and the part of the curve deviation in Eq.~\eqref{eqn:DeltaK_moment}---has a previously understood decomposition into charge and flux parts, because of its relation to the supermomentum charge conjugate to BMS supertranslation symmetries.
We review this decomposition using the Einstein equations rather than the Hamiltonian charges, since the former approach will apply to all moments of the news, whereas it is not yet known if the charge viewpoint would apply to all the moments of the news.\footnote{There is a relationship between the first moment of the news and charges (and fluxes) of the extended~\cite{Barnich2009} or generalized~\cite{Campiglia2014} BMS algebras, which have been computed in Bondi coordinates in the physical spacetime~\cite{Pasterski2015, Compere2019} (see also~\cite{Compere2018, Compere2020}).
  It does not seem possible to obtain these charges through a covariant procedure at null infinity in the unphysical spacetime~\cite{Flanagan2019b}.}

We start with the conservation equation for $m$, Eq.~\eqref{eqn:dot_m}, and we note that the right-hand side vanishes when there is no news.
Integrating this equation, we can solve for the zeroth moment of $N_{ij}$:
\begin{equation} \label{eqn:N0_bondi}
  \frac{1}{4} \ms D^i \ms D^j \ord{0}{\mc N}_{ij} (u_1, u_0) = \Delta m(u_1, u_0) - \int_{u_0}^{u_1} \ud u_2 \mc F (u_2),
\end{equation}
where
\begin{equation}
  \Delta m(u_1, u_0) \equiv m(u_1) - m(u_0), \quad \mc F \equiv -\frac{1}{8} N_{ij} N^{ij}.
\end{equation}
We call the $\Delta m(u_1, u_0)$ term the \emph{charge} contribution, because it is the difference between the values of a ``conserved quantity'' (in this case, $m$) at $u_0$ and $u_1$.
The mass aspect $m$ is a ``conserved quantity'' in the sense that it is conserved in the absence of radiation.
We call the integral of $\mc F$ in Eq.~\eqref{eqn:N0_bondi} the \emph{flux} contribution, because it is a nonlinear quantity that vanishes in the absence of radiation (two properties that would be expected of, for example, a flux of energy).

In the remaining subsections, we derive equations, analogous to Eq.~\eqref{eqn:N0_bondi}, for \emph{all} moments of the news.
For convenience, we summarize the form of these results here, but we leave the precise definitions to the later subsections.
We begin with the first moment, which we find is related to the charge and flux contributions as follows:
\begin{equation} \label{eqn:N1_bondi}
  \begin{split}
    \frac{1}{2} \ms D_k &\STF (\ms D_i \ms D_j) \ord{1}{\mc N}{}^{jk} (u_1, u_0) \\
    &= \int_{u_0}^{u_1} \ud u_2 \tilde{\mc F}_i (u_2, u_0) - \Delta \tilde{N}_i (u_1, u_0).
  \end{split}
\end{equation}
Here, we have defined
\begin{equation}
  \Delta \tilde{N}_i (u_1, u_0) \equiv \tilde{N}_i (u_1, u_0) - \tilde{N}_i (u_0, u_0),
\end{equation}
and $\tilde{N}_i (u, \tilde{u})$ and $\tilde{\mc F}_i (u, \tilde{u})$ are defined in Eqs.~\eqref{eqn:tildeN1} and~\eqref{eqn:tildeF0_i}, respectively.
The quantity $\tilde{N}_i (u, \tilde{u})$, much like $m$, we call a ``conserved quantity,'' because $\partial_u \tilde{N}_i (u, \tilde{u}) = 0$ when the news vanishes.
Note that the angular momentum aspect $N_i$ is \emph{not} a ``conserved quantity'' in this sense, because its $u$ derivative satisfies Eq.~\eqref{eqn:dot_N} and does not vanish when the news is zero.
Additional terms given in Eq.~\eqref{eqn:tildeN1} must be added to $N_i$ to form a quantity, $\tilde{N}_i (u, \tilde{u})$, that has a vanishing $u$ derivative when the news tensor vanishes.
Equation~\eqref{eqn:N1_bondi} has the same form as Eq.~\eqref{eqn:N0_bondi}, with a charge contribution given by $\Delta \tilde{N}_i (u_1, u_0)$ and a flux contribution given by the integral of $\tilde{\mc F}_i (u_2, u_0)$.

Similarly, there are ``corrected'' versions of $\ord{n}{\mc E}_{ij}$, which we denote by $\ord{n}{\tilde{\mc E}}_{ij} (u, \tilde{u})$, such that for any $n \geq 0$, these quantities satisfy $\partial_u \ord{n}{\tilde{\mc E}}_{ij} (u, \tilde{u}) = 0$ when the news vanishes.
Changes in these ``conserved quantities'' are related to the higher moments of the news.
For $n = 0$, we find that the relation is
\begin{equation} \label{eqn:N2_bondi}
  \begin{split}
    \frac{1}{6} \STF &\ms D_i \left[\ms D_l \STF (\ms D_j \ms D_k) \ord{2}{\mc N}^{kl} (u_1, u_0)\right] \\
    &= \Delta \ord{0}{\tilde{\mc E}}_{ij} (u_1, u_0) - \int_{u_0}^{u_1} \ud u_2 \ord{0}{\tilde{\mc F}}_{ij} (u_2, u_0),
  \end{split}
\end{equation}
while for higher moments we find that (for $n \geq 0$)
\begin{equation} \label{eqn:Nn3_bondi}
  \begin{split}
    \frac{(-1)^{n + 1}}{6} \mc D_n &\cdots \mc D_0 \STF \ms D_i \!\left[\ms D_l \STF (\ms D_j \ms D_k) \ord{n + 3}{\mc N}^{kl} (u_1, u_0)\right] \\
    &= \Delta \ord{n + 1}{\tilde{\mc E}}_{ij} (u_1, u_0) - \int_{u_0}^{u_1} \ud u_2 \ord{n + 1}{\tilde{\mc F}}_{ij} (u_2, u_0).
  \end{split}
\end{equation}
In these equations, we have defined
\begin{equation}
  \Delta \ord{n}{\tilde{\mc E}}_{ij} (u_1, u_0) \equiv \ord{n}{\tilde{\mc E}}_{ij} (u_1, u_0) - \ord{n}{\tilde{\mc E}}_{ij} (u_0, u_0),
\end{equation}
where $\ord{0}{\tilde{\mc E}}_{ij} (u, \tilde{u})$ is defined in Eq.~\eqref{eqn:tildeE0} and $\ord{n + 1}{\tilde{\mc E}}_{ij} (u, \tilde{u})$ (for $n \geq 0$) is defined in Eq.~\eqref{eqn:tildeEn1}.
The contribution to each moment of the news due to these terms we will call the charge contribution, as before.
Note that equations similar to Eqs.~\eqref{eqn:N2_bondi} and~\eqref{eqn:Nn3_bondi} are also given, in the linear theory, by Eqs.~(11) and~(12) of \cite{Mao2020}, which are written in terms of components of the Weyl tensor, instead of $\ord{n}{\mc E}_{ij}$.

The flux contributions are similarly given by the integrals of $\ord{n}{\tilde{\mc F}}_{ij} (u_2, u_0)$, but unlike the case of the zeroth and first moments of the news, these contributions can be further divided into two pieces:
\begin{equation}
  \ord{n}{\tilde{\mc F}}_{ij} (u, \tilde{u}) \equiv \ord{n}{\tilde{\mc F}}^{\textrm{rad}}_{ij} (u, \tilde{u}) + \ord{n}{\tilde{\mc F}}^{\textrm{nonrad}}_{ij} (u, \tilde{u}).
\end{equation}
Here $\ord{n}{\tilde{\mc F}}^{\textrm{rad}}_{ij} (u, \tilde{u})$ is a nonlinear function constructed from $C_{ij}$ and $N_{ij}$ (the radiative degrees of freedom), whereas $\ord{n}{\tilde{\mc F}}^{\textrm{nonrad}}_{ij} (u, \tilde{u})$ depends also on at least one of the nonradiative degrees of freedom in $m$, $N_i$, and $\ord{p}{\mc E}_{ij}$ (for any $p < n$).
Equation~\eqref{eqn:tildeF0} contains the values of $\ord{0}{\tilde{\mc F}}^{\textrm{rad}}_{ij} (u, \tilde{u})$ and $\ord{0}{\tilde{\mc F}}^{\textrm{nonrad}}_{ij} (u, \tilde{u})$, and Eq.~\eqref{eqn:tildeFn1} contains the values of $\ord{n + 1}{\tilde{\mc F}}^{\textrm{rad}}_{ij} (u, \tilde{u})$ and $\ord{n + 1}{\tilde{\mc F}}^{\textrm{nonrad}}_{ij} (u, \tilde{u})$ (for $n \geq 0$).

The existence of these nonradiative flux contributions are unique to the second and higher moments of the news, as the flux contributions to the zeroth and first moments only contain $C_{ij}$ and $N_{ij}$.
As before, all of these flux contributions vanish when the news vanishes.

In the remaining subsections, we give the derivations of Eq.~\eqref{eqn:N1_bondi} in Sec.~\ref{sec:N1}, of Eq.~\eqref{eqn:N2_bondi} in Sec.~\ref{sec:N2}, and of Eq.~\eqref{eqn:Nn3_bondi} in Sec.~\ref{sec:Nn3}.
We then conclude this section with a discussion of solving Eqs.~\eqref{eqn:N0_bondi}, \eqref{eqn:N1_bondi}, \eqref{eqn:N2_bondi}, and~\eqref{eqn:Nn3_bondi} for these temporal moments of the news using an expansion of these quantities in tensor harmonics in Sec.~\ref{sec:harmonics}.

\subsection{Derivations of the moments}

\subsubsection{First moment} \label{sec:N1}

We start with the first moment of the news; qualitatively, the calculations and procedures that we use here will carry over to the calculations for higher moments.
Consider the conservation equation for the angular momentum aspect, $N_i$, in Eq.~\eqref{eqn:dot_N}.
To extract from this expression an equation like Eq.~\eqref{eqn:N0_bondi}, we note that, unlike the conservation equation for $m$, the right-hand side of Eq.~\eqref{eqn:dot_N} has two types of terms: those that vanish when news vanishes, and those that do not.
Consequently, $N_i (u_1) - N_i (u_0)$ does \emph{not} vanish when there is no news between $u_0$ and $u_1$, unlike $\Delta m(u_1, u_0)$.\footnote{A closely related statement is that the mass aspect $m$ is independent of $u$ in regions without news, whereas the angular momentum aspect $N_i$ depends linearly on $u$ (for example, see~\cite{Bondi1962, Flanagan2015}).}

It is possible to construct from $N_i$ a new quantity, $\tilde{N}_i (u, \tilde{u})$, which has the property that $\partial_u \tilde{N}_i (u, \tilde{u}) = 0$ when the news is zero.
First, note that $\tilde{N}_i (u, \tilde{u})$ depends on a ``reference'' time $\tilde{u}$.
Second, note that this construction is not unique: related quantities enter into the charges defined by Wald and Zoupas~\cite{Wald1999, Flanagan2015}, the two-parameter family of charges in~\cite{Compere2019, Elhashash2021}, and the supertranslation-invariant angular momentum in~\cite{Chen2021}.
The particular choice that we make here is just a simple example of such a quantity:
\begin{equation} \label{eqn:tildeN1}
  \tilde{N}_i (u, \tilde{u}) \equiv N_i (u) - (u - \tilde{u}) \ms D^j m_{ij} (u),
\end{equation}
where
\begin{equation}
  m_{ij} \equiv m h_{ij} + \frac{1}{2} \ms D_{[i} \ms D^k C_{j]k}.
\end{equation}

To show that $\partial_u \tilde{N}_i (u, \tilde{u})$ vanishes when there is no news, we first point out that $\partial_u m_{ij}$ is zero in the absence of news:
\begin{equation}
  \partial_u m_{ij} = \frac{1}{2} \STF (\ms D_i \ms D_k) N^k{}_j + \mc F h_{ij}.
\end{equation}
To compute this expression, we used the fact that, for any tensor operator $A_{ij}$,
\begin{equation}
  \begin{split}
    A_{[i}{}^k N_{j]k} + \frac{1}{2} A_{kl} N^{kl} h_{ij} &= A_i{}^k N_{jk} - \STF (A_i{}^k N_{jk}) \\
    &= \STF (A_{ik}) N^k{}_j.
  \end{split}
\end{equation}
The second line follows from the fact that $N_{ij}$ is a symmetric, trace-free tensor.
Finally, note that the terms in Eq.~\eqref{eqn:dot_N} that are nonzero when the news is nonzero are given by $\ms D^j m_{ij}$; however, the $u$ derivative of the second term of Eq.~\eqref{eqn:tildeN1} cancels this term in regions with a vanishing news tensor.

To summarize, we find that
\begin{equation} \label{eqn:tildeN1_dot}
  \begin{split}
    \partial_u \tilde{N}_i (u, \tilde{u}) &= \tilde{\mc F}_i (u, \tilde{u}) - \frac{1}{2} (u - \tilde{u}) \ms D_k \STF (\ms D_i \ms D_j) N^{jk} (u),
  \end{split}
\end{equation}
where
\begin{equation} \label{eqn:tildeF0_i}
  \tilde{\mc F}_i (u, \tilde{u}) \equiv \mc F_i (u) - (u - \tilde{u}) \ms D_i \mc F (u),
\end{equation}
and where
\begin{equation}
  \mc F_i \equiv \frac{1}{4} (N^{jk} \ms D_j C_{ki} + 3 C_{ij} \ms D_k N^{jk}).
\end{equation}
Integrating Eq.~\eqref{eqn:tildeN1_dot} from $u_0$ to $u_1$, and setting $\tilde{u} = u_0$, produces Eq.~\eqref{eqn:N1_bondi}.

\subsubsection{Second moment} \label{sec:N2}

We now relate the second moment of the news to the evolution equation for $\ord{0}{\mc E}_{ij}$ in Eq.~\eqref{eqn:dot_E0}.
Many terms in~\eqref{eqn:dot_E0} are nonzero when the news vanishes; thus, to make a charge and flux decomposition, we must define a quantity $\ord{0}{\tilde{\mc E}}_{ij}$ from $\ord{0}{\mc E}_{ij}$ that has vanishing $u$ derivative when the news tensor is zero.
We first write this evolution equation in the following form, which will be similar to the form that the evolution equations take at higher order in $1/r$:
\begin{equation} \label{eqn:partial_u_Eij0}
  \partial_u \ord{0}{\mc E}_{ij} = \ord{0}{\mc F}^{\textrm{rad}}_{ij} + \frac{1}{3} \STF \ms D_i N_j + \ord{0, 0}{\mc G}^{\textrm{rad}}_{ij} + \ord{0, 0}{\mc G}^{\textrm{nonrad}}_{ij}.
\end{equation}
Here, we have defined
\begin{equation}
  \ord{0}{\mc F}^{\textrm{rad}}_{ij} \equiv \frac{1}{4} N_{kl} C^{kl} C_{ij}
\end{equation}
to be the nonlinear piece of this equation that depends on radiative degrees of freedom and vanishes when there is no news (note that it is cubic, rather than quadratic in the news and shear),
\begin{equation}
  \ord{0, 0}{\mc G}^{\textrm{rad}}_{ij} \equiv \frac{1}{4} C^k{}_j \ms D_{[i} \ms D^l C_{k]l}
\end{equation}
to be a piece that does not vanish when there is no news, but is a nonlinear function of the shear, and
\begin{equation}
  \ord{0, 0}{\mc G}^{\textrm{nonrad}}_{ij} \equiv \frac{1}{2} m C_{ij}
\end{equation}
to be the nonlinear piece that depends on a least one of the nonradiative degrees of freedom.
The notation with two numbers underset below these tensors will be explained in the next subsection.

Using this decomposition of the evolution equation, we now define the quantity $\ord{0}{\mc E}_{ij}$ by
\begin{equation} \label{eqn:tildeE0}
  \begin{split}
    \ord{0}{\tilde{\mc E}}_{ij} (u, \tilde{u}) \equiv \ord{0}{\mc E}_{ij} (u) + (\tilde{u} - u) \bigg\{&\frac{1}{3} \STF \ms D_i \ord{2}{\tilde N}_j (u, \tilde{u}) \\
    &+ \ord{0, 0}{\mc G}^{\textrm{rad}}_{ij} (u) \\
    &+ \ord{0, 0}{\mc G}^{\textrm{nonrad}}_{ij} (u)\bigg\},
  \end{split}
\end{equation}
where
\begin{equation}
  \ord{n}{\tilde N}_i (u, \tilde{u}) \equiv N_i (u) + \frac{\tilde{u} - u}{n} \ms D^j m_{ij} (u)
\end{equation}
[note that $\tilde{N}_i (u, \tilde{u})$ defined in the previous section is given by $\ord{1}{\tilde N}_i (u, \tilde{u})$].
The $u$ derivative of $\ord{n}{\tilde N}_i (u, \tilde{u}) $ satisfies a generalization of Eq.~\eqref{eqn:tildeN1_dot}, which is given by
\begin{equation} \label{eqn:tildeNn_dot}
  \begin{split}
    \partial_u \bigg[&\frac{(\tilde{u} - u)^{n + 1}}{(n + 1)!} \ord{n + 2}{\tilde N}_i (u, \tilde{u})\bigg] \\
    &= -\frac{(\tilde{u} - u)^n}{n!} N_i (u) + \frac{(\tilde{u} - u)^{n + 1}}{(n + 1)!} \ord{n + 1}{\tilde{\mc F}}_i (u, \tilde{u}) \\
    &\hspace{1em}+ \frac{(\tilde{u} - u)^{n + 2}}{2 (n + 2)!} \ms D_k \STF (\ms D_i \ms D_j) N^{jk} (u),
  \end{split}
\end{equation}
where
\begin{equation}
  \ord{n}{\tilde{\mc F}}_i (u, \tilde{u}) \equiv \left[\mc F_i (u) + \frac{(\tilde{u} - u)}{n + 1} \ms D_i \mc F(u)\right]
\end{equation}
[so that $\tilde{\mc F}_i (u, \tilde{u})$, as defined in Eq.~\eqref{eqn:tildeF0_i}, is simply $\ord{0}{\tilde{\mc F}}_i (u, \tilde{u})$].

Next let us take the derivative of $\ord{0, 0}{\mc G}^{\textrm{nonrad}}_{ij} (u)$, which we write as
\begin{equation} \label{eqn:G00_dot}
  \partial_u \ord{0, 0}{\mc G}^{\textrm{nonrad}}_{ij} = \ord{0, 1}{\mc F}^{\textrm{rad}}_{ij} + \ord{0, 1}{\mc F}^{\textrm{nonrad}}_{ij},
\end{equation}
where
\begin{equation}
  \ord{0, 1}{\mc F}^{\textrm{rad}}_{ij} \equiv \frac{1}{8} \left(\ms D_k \ms D_l N^{kl} - \frac{1}{2} N_{kl} N^{kl}\right) C_{ij}
\end{equation}
is the piece of Eq.~\eqref{eqn:G00_dot} that depends only on radiative degrees of freedom and
\begin{equation}
  \ord{0, 1}{\mc F}^{\textrm{nonrad}}_{ij} \equiv \frac{1}{2} m N_{ij}
\end{equation}
is the piece of Eq.~\eqref{eqn:G00_dot} that depends on nonradiative degrees of freedom.
Note that the right-hand side of Eq.~\eqref{eqn:G00_dot} vanishes when the news vanishes [that is, $\ord{0, 0}{\mc G}^{\textrm{nonrad}}_{ij} (u)$ has vanishing $u$ derivative].

We now have all the necessary elements to compute the $u$ derivative of $\ord{0}{\tilde{\mc E}}_{ij} (u, \tilde{u})$, which will give us the charge and flux decomposition of the second moment of the news.
We then find that [using Eqs.~\eqref{eqn:tildeE0}, \eqref{eqn:partial_u_Eij0}, and~\eqref{eqn:G00_dot}, as well as Eq.~\eqref{eqn:tildeNn_dot} for $n = 0$]
\begin{equation} \label{eqn:tildeE0dot}
  \begin{split}
    \partial_u &\ord{0}{\tilde{\mc E}}_{ij} (u, \tilde{u}) \\
    &= \ord{0}{\tilde{\mc F}}^{\textrm{rad}}_{ij} (u, \tilde{u}) + \ord{0}{\tilde{\mc F}}^{\textrm{nonrad}}_{ij} (u, \tilde{u}) \\
    &\hspace{1em}+ \frac{(\tilde{u} - u)^2}{12} \STF \ms D_i \left[\ms D_l \STF (\ms D_j \ms D_k) N^{kl} (u)\right],
  \end{split}
\end{equation}
where
\begin{subequations} \label{eqn:tildeF0}
  \begin{align}
    \ord{0}{\tilde{\mc F}}^{\textrm{rad}}_{ij} (u, \tilde{u}) &\equiv \ord{0}{\mc F}^{\textrm{rad}}_{ij} (u) \nonumber \\
    &\hspace{1em}+ (\tilde{u} - u) \bigg[\frac{1}{3} \STF \ms D_i \ord{1}{\tilde{\mc F}}_j (u, \tilde{u}) \nonumber \\
    &\hspace{6em}+ \ord{0, 0}{\dot{\mc G}}^{\textrm{rad}}_{ij} (u) + \ord{0, 1}{\mc F}^{\textrm{rad}}_{ij} (u)\bigg], \\
    \ord{0}{\tilde{\mc F}}^{\textrm{nonrad}}_{ij} (u, \tilde{u}) &\equiv (\tilde{u} - u) \ord{0, 1}{\mc F}^{\textrm{nonrad}}_{ij} (u).
  \end{align}
\end{subequations}
Equation~\eqref{eqn:tildeE0dot} can now be integrated in time to yield Eq.~\eqref{eqn:N2_bondi} for the second moment of the news, upon setting $\tilde{u} = u_0$.

\subsubsection{Procedure for higher moments} \label{sec:Nn3}

The procedure required to compute the charge and flux contributions to the order $n + 3$ moment of the news is similar to that used in Sec.~\ref{sec:N2} to determine the second moment.
It first requires the evolution equation for $\ord{n + 1}{\mc E}_{ij}$, for $n \geq 0$.
Given the linearized limit of this evolution equation in Eq.~\eqref{eqn:dot_En}, and using the full nonlinear expression that can be obtained by Eq.~\eqref{eqn:dot_C}, the evolution equation must take the following form:
\begin{equation} \label{eqn:general_evol}
  \begin{split}
    \partial_u \ord{n + 1}{\mc E}_{ij} &= \ord{n + 1}{\mc F}^{\textrm{rad}}_{ij} + \ord{n + 1}{\mc F}^{\textrm{nonrad}}_{ij} \\
    &\hspace{1em}+ \mc D_n \ord{n}{\mc E}_{ij} + \ord{n + 1, 0}{\mc G}^{\textrm{rad}}_{ij} + \ord{n + 1, 0}{\mc G}^{\textrm{nonrad}}_{ij}.
  \end{split}
\end{equation}
While we do not compute the exact expressions for these quantities in terms of the Bondi metric functions, the properties of the four nonlinear terms can be summarized as follows:
\begin{enumerate}[(i)]

\item $\ord{n + 1}{\mc F}^{\textrm{rad}}_{ij}$ depends nonlinearly on $C_{ij}$ and $N_{ij}$ and each term in its expression must include one copy of $N_{ij}$;

\item $\ord{n + 1}{\mc F}^{\textrm{nonrad}}_{ij}$ also depends nonlinearly on $C_{ij}$ and $N_{ij}$ (similarly to $\ord{n+1}{\mc F}_{ij}$), but in addition has dependence on at least one of $m$, $N^i$, and $\ord{p}{\mc E}_{ij}$ (for $p < n + 1$);

\item $\ord{n + 1, 0}{\mc G}^{\textrm{rad}}_{ij}$ depends nonlinearly on $C_{ij}$ alone; and

\item $\ord{n + 1, 0}{\mc G}^{\textrm{nonrad}}_{ij}$ depends nonlinearly on $C_{ij}$, as well as on at least one of $m$, $N^i$, and $\ord{p}{\mc E}_{ij}$ (for $p < n$).

\end{enumerate}
This decomposition can be determined from the full nonlinear expression in Eq.~\eqref{eqn:dot_C} by noting that on the right-hand side for expression for $\partial_u \ord{n + 1}{\mc E}_{ij}$ there are no terms with explicit $u$ derivatives (the only term with an implicit $u$ derivative is $N_{ij} = \partial_u C_{ij}$, which appears at most once in each term on the right-hand side).
The terms $\ord{n + 1}{\mc F}^{\textrm{rad}}_{ij}$ and $\ord{n + 1}{\mc F}^{\textrm{nonrad}}_{ij}$ are the only ones which vanish when the news vanishes, and $\mc D_n \ord{n}{\mc E}_{ij}$ is the only term in this equation that is linear.
Note that the constraints on the values of $p$ that can occur in the dependence of $\ord{n + 1}{\mc F}^{\textrm{nonrad}}_{ij}$ and $\ord{n + 1, 0}{\mc G}^{\textrm{nonrad}}_{ij}$ on $\ord{p}{\mc E}_{ij}$ can be determined by dimensional analysis.

Constructing a quantity  $\ord{n + 1}{\tilde{\mc E}}_{ij}$ from $ \ord{n + 1}{\mc E}_{ij}$ becomes more involved than it was for $\ord{0}{\mc E}_{ij}$, because unlike $\partial_u \ord{0, 0}{\mc G}^{\textrm{nonrad}}_{ij}$, the quantity $\partial_u \ord{n + 1, 0}{\mc G}^{\textrm{nonrad}}_{ij}$ does not generally vanish when the news is zero.
However, we can construct the following iterative procedure to find a quantity related to  higher $u$ derivatives of $\ord{n + 1, 0}{\mc G}^{\textrm{nonrad}}_{ij}$ that does eventually have a $u$ derivative that vanishes when the news tensor is zero.
To obtain this quantity, consider the iterative relationship:
\begin{equation}
  \begin{split}
    \partial_u \ord{n, q}{\mc G}_{ij}^{\textrm{nonrad}} &\equiv \ord{n, q + 1}{\mc F}^{\textrm{rad}}_{ij} + \ord{n, q + 1}{\mc F}^{\textrm{nonrad}}_{ij} \\
    &\hspace{1em}+ \ord{n, q + 1}{\mc G}^{\textrm{rad}}_{ij} + \ord{n, q + 1}{\mc G}^{\textrm{nonrad}}_{ij},
  \end{split}
\end{equation}
where
\begin{enumerate}[(i)]

\item $\ord{n, q + 1}{\mc F}^{\textrm{rad}}_{ij}$ depends nonlinearly on $C_{ij}$ and $N_{ij}$;

\item $\ord{n, q + 1}{\mc F}^{\textrm{nonrad}}_{ij}$ depends nonlinearly on $C_{ij}$, $N_{ij}$, along with at least one of $m$, $N^i$, and $\ord{p}{\mc E}_{ij}$ (for $p < n - q - 1$);

\item $\ord{n, q + 1}{\mc G}^{\textrm{rad}}_{ij}$ depends nonlinearly on $C_{ij}$ alone; and

\item $\ord{n, q + 1}{\mc G}^{\textrm{nonrad}}_{ij}$ depends nonlinearly on $C_{ij}$, as well as on at least one of $m$, $N^i$, and $\ord{p}{\mc E}_{ij}$ (for $p < n - q - 2$).

\end{enumerate}
Again the values of $p$ that are allowed are constrained by dimensional analysis.
Note that this process of iteratively defining $\ord{n, q}{\mc G}^{\textrm{nonrad}}_{ij}$ will end at some point, as there always exists a $q_n$ such that
\begin{equation}
  \ord{n, q_n}{\mc G}^{\textrm{nonrad}}_{ij} = 0.
\end{equation}
In particular, from Eq.~\eqref{eqn:G00_dot}, we have that $q_0 = 1$.

To define $\ord{n + 1}{\tilde{\mc E}}_{ij} (u, \tilde{u})$ from $\ord{n + 1}{\mc E}_{ij}$, we need to introduce a few more quantities constructed from $\ord{n, q}{\mc G}^{\textrm{rad}}_{ij}$ and $\ord{n, q}{\mc G}^{\textrm{nonrad}}_{ij}$.
These quantities are
\begin{widetext}
\begin{equation} \label{eqn:tildeG}
  \ord{n, p}{\tilde{\mc G}}_{ij} (u, \tilde{u}) \equiv \sum_{q = 0}^{q_{n - p}} \frac{(p + 2)! (\tilde{u} - u)^q}{(p + q + 2)!} \left[\ord{n - p, q}{\mc G}^{\textrm{rad}}_{ij} (u) + \ord{n - p, q}{\mc G}^{\textrm{nonrad}}_{ij} (u)\right],
\end{equation}
which has the property that
\begin{equation}
  \begin{split}
    \partial_u \left[\frac{(\tilde{u} - u)^{p + 2}}{(p + 2)!} \ord{n, p}{\tilde{\mc G}}_{ij} (u, \tilde{u})\right] &= -\frac{(\tilde{u} - u)^{p + 1}}{(p + 1)!} \left[\ord{n - p, 0}{\mc G}^{\textrm{rad}}_{ij} (u) + \ord{n - p, 0}{\mc G}^{\textrm{nonrad}}_{ij} (u)\right] \\
    &\hspace{1em}+ \frac{(\tilde{u} - u)^{p + 2}}{(p + 2)!} \left[\ord{n, p}{\tilde{\mc F}}^{\textrm{rad}} (u, \tilde{u}) + \ord{n, p}{\tilde{\mc F}}^{\textrm{nonrad}} (u, \tilde{u})\right],
  \end{split}
\end{equation}
where
\begin{subequations}
  \begin{align}
    \ord{n, p}{\tilde{\mc F}}^{\textrm{rad}}_{ij} (u, \tilde{u}) &\equiv \sum_{q = 0}^{q_{n - p} - 1} \frac{(p + 2)! (\tilde{u} - u)^q}{(p + q + 2)!} \left[\ord{n - p, q + 1}{\mc F}^{\textrm{rad}}_{ij} (u) + \ord{n - p, q}{\dot{\mc G}}^{\textrm{rad}}_{ij} (u)\right] + \frac{(p + 2)! (\tilde{u} - u)^{q_{n - p}}}{(p + q_{n - p} + 2)!} \ord{n - p, q_{n - p}}{\dot{\mc G}}^{\textrm{rad}}_{ij} (u), \label{eqn:tildeFnp_rad} \\
    \ord{n, p}{\tilde{\mc F}}^{\textrm{nonrad}}_{ij} (u, \tilde{u}) &\equiv \sum_{q = 0}^{q_{n - p} - 1} \frac{(p + 2)! (\tilde{u} - u)^q}{(p + q + 2)!} \ord{n - p, q + 1}{\mc F}^{\textrm{nonrad}}_{ij} (u). \label{eqn:tildeFnp_nonrad}
  \end{align}
\end{subequations}
The definition of $\ord{n, p}{\tilde{\mc G}}_{ij} (u, \tilde{u})$ was created so that it cancels terms that appear in the evolution equation for $\ord{n - p}{\mc E}_{ij}$ that do not vanish when the news is zero.
Using this definition, it is then possible to show that the following definition
\begin{equation} \label{eqn:tildeEn1}
  \begin{split}
    \ord{n + 1}{\tilde{\mc E}}_{ij} (u, \tilde{u}) \equiv \ord{n + 1}{\mc E}_{ij} (u) &+ (\tilde{u} - u) \ord{n, -1}{\tilde{\mc G}}_{ij} (u, \tilde{u}) + \sum_{p = 0}^n \frac{(\tilde{u} - u)^{p + 1}}{(p + 1)!} \mc D_n \cdots \mc D_{n - p} \left[\ord{n - p}{\mc E}_{ij} (u) + \frac{\tilde{u} - u}{p + 2} \ord{n, p}{\tilde{\mc G}}_{ij} (u, \tilde{u})\right] \\
    &+ \frac{(\tilde{u} - u)^{n + 2}}{3 (n + 2)!} \mc D_n \cdots \mc D_0 \STF \ms D_i \ord{n + 3}{\tilde N}_j (u, \tilde{u}),
  \end{split}
\end{equation}
satisfies
\begin{equation} \label{eqn:tildeEn1_dot}
  \partial_u \ord{n + 1}{\tilde{\mc E}}_{ij} (u, \tilde{u}) = \ord{n + 1}{\tilde{\mc F}}^{\textrm{rad}}_{ij} (u, \tilde{u}) + \ord{n + 1}{\tilde{\mc F}}^{\textrm{nonrad}}_{ij} (u, \tilde{u}) + \frac{(\tilde{u} - u)^{n + 3}}{6 (n + 3)!} \mc D_n \cdots \mc D_0 \STF \ms D_i \left[\ms D_l \STF (\ms D_j \ms D_k) N^{kl}\right],
\end{equation}
where
\begin{subequations} \label{eqn:tildeFn1}
  \begin{align}
    \ord{n + 1}{\tilde{\mc F}}^{\textrm{rad}}_{ij} (u, \tilde{u}) &\equiv \ord{n + 1}{\mc F}^{\textrm{rad}}_{ij} (u) + (\tilde{u} - u) \ord{n, -1}{\tilde{\mc F}}^{\textrm{rad}}_{ij} (u, \tilde{u}) + \sum_{p = 0}^n \frac{(\tilde{u} - u)^{p + 1}}{(p + 1)!} \mc D_n \cdots \mc D_{n - p} \left[\ord{n - p}{\mc F}^{\textrm{rad}}_{ij} (u) + \frac{\tilde{u} - u}{p + 2} \ord{n, p}{\tilde{\mc F}}^{\textrm{rad}}_{ij} (u, \tilde{u})\right] \nonumber \\
    &\hspace{1em}+ \frac{(\tilde{u} - u)^{n + 2}}{3 (n + 2)!} \mc D_n \cdots \mc D_0 \STF \ms D_i \ord{n + 2}{\tilde{\mc F}}_j (u, \tilde{u}), \\
    \ord{n + 1}{\tilde{\mc F}}^{\textrm{nonrad}}_{ij} (u, \tilde{u}) &\equiv \ord{n + 1}{\mc F}^{\textrm{nonrad}}_{ij} (u) + (\tilde{u} - u) \ord{n, -1}{\tilde{\mc F}}^{\textrm{nonrad}}_{ij} (u, \tilde{u}) \nonumber \\
    &\hspace{1em}+ \sum_{p = 0}^n \frac{(\tilde{u} - u)^{p + 1}}{(p + 1)!} \mc D_n \cdots \mc D_{n - p} \left[\ord{n - p}{\mc F}^{\textrm{nonrad}}_{ij} (u) + \frac{\tilde{u} - u}{p + 2} \ord{n, p}{\tilde{\mc F}}^{\textrm{nonrad}}_{ij} (u, \tilde{u})\right].
  \end{align}
\end{subequations}
Thus, $\partial_u \ord{n + 1}{\tilde{\mc E}}_{ij} (u, \tilde{u})$ vanishes when there is vanishing news.
Equation~\eqref{eqn:tildeEn1_dot} then can be integrated in time to yield the charge and flux decomposition for the order $n + 3$ moment of the news given in Eq.~\eqref{eqn:Nn3_bondi}.

\end{widetext}

\subsubsection{Procedure applied to the third moment}

As the procedure in Sec.~\ref{sec:Nn3} for classifying the terms in the evolution equation and constructing the charge and flux decomposition for the moments of the news is somewhat involved, we provide an example for $n = 0$, since the evolution equation for $\ord{1}{\mc E}_{ij}$ is known [and given by Eq.~\eqref{eqn:dot_E1}].
Inspecting the various terms in Eq.~\eqref{eqn:dot_E1}, we see that
\begin{equation}
  \ord{1}{\mc F}^{\textrm{rad}}_{ij} = \ord{1}{\mc F}^{\textrm{nonrad}}_{ij} = 0,
\end{equation}
along with
\begin{subequations}
  \begin{align}
    \ord{1, 0}{\mc G}^{\textrm{rad}}_{ij} &= \frac{1}{4} \STF \ms D^k \bigg\{\bigg[C_{il} \ms D_m C^{lm} \nonumber \\
    &\hspace{6.8em}- \frac{3}{8} \ms D_i (C_{lm} C^{lm})\bigg] C_{jk} \nonumber \\
    &\hspace{6.4em}+ \frac{5}{8} C_{lm} C^{lm} \ms D_i C_{jk}\bigg\}, \label{eqn:10_G_rad} \\
    \ord{1, 0}{\mc G}^{\textrm{nonrad}}_{ij} &= -\frac{1}{3} \STF \ms D^k (N_i C_{jk}). \label{eqn:10_G_nonrad}
  \end{align}
\end{subequations}
Taking a $u$ derivative of Eq.~\eqref{eqn:10_G_nonrad}, we find that
\begin{subequations}
  \begin{align}
    \ord{1, 1}{\mc F}^{\textrm{rad}}_{ij} &= -\frac{1}{12} \STF \ms D^k [(N^{lm} \ms D_l C_{mi} \nonumber \\
    &\hspace{7.5em}+ 3 C_{il} \ms D_m N^{lm}) C_{jk}], \\
    \ord{1, 1}{\mc F}^{\textrm{nonrad}}_{ij} &= -\frac{1}{3} \STF \ms D^k (N_i N_{jk}) \displaybreak[0] \\
    \ord{1, 1}{\mc G}^{\textrm{rad}}_{ij} &= \frac{1}{6} \STF \ms D^k (C_{jk}\ms D^l \ms D_{[l} \ms D^m C_{i]m}), \label{eqn:11_G_rad} \\
    \ord{1, 1}{\mc G}^{\textrm{nonrad}}_{ij} &= -\frac{1}{3} \STF \ms D^k (C_{jk} \ms D_i m), \label{eqn:11_G_nonrad}
  \end{align}
\end{subequations}
and taking a $u$ derivative of Eq.~\eqref{eqn:11_G_nonrad}, we find that
\begin{subequations}
  \begin{align}
    \ord{1, 2}{\mc F}^{\textrm{rad}}_{ij} &= -\frac{1}{12} \STF \ms D^k \bigg[C_{jk} \ms D_i \bigg(\ms D_l \ms D_m N^{lm} \nonumber \\
    &\hspace{10.7em}- \frac{1}{2} N_{lm} N^{lm}\bigg)\bigg], \\
    \ord{1, 2}{\mc F}^{\textrm{nonrad}}_{ij} &= -\frac{1}{3} \STF \ms D^k (N_{jk} \ms D_i m),
  \end{align}
\end{subequations}
along with
\begin{equation}
  \ord{1, 2}{\mc G}^{\textrm{rad}}_{ij} = \ord{1, 2}{\mc G}^{\textrm{nonrad}}_{ij} = 0,
\end{equation}
so that $q_1 = 2$.

With these quantities defined, we can now write down a more explicit expression for $\ord{1}{\tilde{\mc E}}_{ij} (u, \tilde{u})$.
By Eq.~\eqref{eqn:tildeEn1}, we have that
\begin{equation}
  \begin{split}
    \ord{1}{\tilde{\mc E}}_{ij} (u, \tilde{u}) &= \ord{1}{\mc E}_{ij} (u) + (\tilde{u} - u) \ord{0, -1}{\tilde{\mc G}}_{ij} (u, \tilde{u}) \\
    &\hspace{1em}+ (\tilde{u} - u) \mc D_0 \left[\ord{0}{\mc E}_{ij} (u) + \frac{\tilde{u} - u}{2} \ord{0, 0}{\tilde{\mc G}}_{ij} (u, \tilde{u})\right] \\
    &\hspace{1em}+ \frac{(\tilde{u} - u)^2}{6} \mc D_0 \STF \ms D_i \ord{3}{\tilde N}_j (u, \tilde{u}),
  \end{split}
\end{equation}
where Eq.~\eqref{eqn:tildeG} implies that
\begin{subequations}
  \begin{align}
    \ord{0, -1}{\tilde{\mc G}}_{ij} (u, \tilde{u}) &= \ord{1, 0}{\mc G}^{\textrm{rad}}_{ij} (u) + \ord{1, 0}{\mc G}^{\textrm{nonrad}}_{ij} (u) \nonumber \\
    &\hspace{1em}+ \frac{\tilde{u} - u}{2} \left[\ord{1, 1}{\mc G}^{\textrm{rad}}_{ij} (u) + \ord{1, 1}{\mc G}^{\textrm{nonrad}}_{ij} (u)\right], \\
    \ord{0, 0}{\tilde{\mc G}}_{ij} (u, \tilde{u}) &= \ord{0, 0}{\mc G}^{\textrm{rad}}_{ij} (u) + \ord{0, 0}{\mc G}^{\textrm{nonrad}}_{ij} (u).
  \end{align}
\end{subequations}
Finally, we can write down more explicit expressions for the radiative and nonradiative flux contributions defined in Eq.~\eqref{eqn:tildeFn1}:
\begin{subequations}
  \begin{align}
    \ord{1}{\tilde{\mc F}}^{\textrm{rad}}_{ij} (u, \tilde{u}) &= (\tilde{u} - u) \ord{0, -1}{\tilde{\mc F}}^{\textrm{rad}}_{ij} (u, \tilde{u}) \nonumber \\
    &\hspace{1em}+ (\tilde{u} - u) \mc D_0 \bigg[\ord{0}{\mc F}^{\textrm{rad}}_{ij} (u) \nonumber \\
    &\hspace{7em}+ \frac{\tilde{u} - u}{2} \ord{0, 0}{\tilde{\mc F}}^{\textrm{rad}}_{ij} (u, \tilde{u})\bigg] \nonumber \\
    &\hspace{1em}+ \frac{(\tilde{u} - u)^2}{6} \mc D_0 \STF \ms D_i \ord{2}{\tilde{\mc F}}_j (u, \tilde{u}), \\
    \ord{1}{\tilde{\mc F}}^{\textrm{nonrad}}_{ij} (u, \tilde{u}) &= (\tilde{u} - u) \ord{0, -1}{\tilde{\mc F}}^{\textrm{nonrad}}_{ij} (u, \tilde{u}) \nonumber \\
    &\hspace{1em}+ (\tilde{u} - u) \mc D_0 \bigg[\ord{0}{\mc F}^{\textrm{nonrad}}_{ij} (u) \nonumber \\
    &\hspace{7em}+ \frac{\tilde{u} - u}{2} \ord{0, 0}{\tilde{\mc F}}^{\textrm{nonrad}}_{ij} (u, \tilde{u})\bigg],
  \end{align}
\end{subequations}
where~\eqref{eqn:tildeFnp_rad} implies that
\begin{subequations}
  \begin{align}
    \ord{0, -1}{\tilde{\mc F}}^{\textrm{rad}}_{ij} (u, \tilde{u}) &= \ord{1, 1}{\mc F}^{\textrm{rad}}_{ij} (u) + \ord{1, 0}{\dot{\mc G}}^{\textrm{rad}}_{ij} (u) \nonumber \\
    &\hspace{1em} + \frac{\tilde{u} - u}{2} \left[\ord{1, 2}{\mc F}^{\textrm{rad}}_{ij} (u) + \ord{1, 1}{\dot{\mc G}}^{\textrm{rad}}_{ij} (u)\right], \\
    \ord{0, 0}{\tilde{\mc F}}^{\textrm{rad}}_{ij} (u, \tilde{u}) &= \ord{0, 1}{\mc F}^{\textrm{rad}}_{ij} (u) + \ord{0, 0}{\dot{\mc G}}^{\textrm{rad}}_{ij} (u),
  \end{align}
\end{subequations}
and Eq.~\eqref{eqn:tildeFnp_nonrad} implies that
\begin{subequations}
  \begin{align}
    \ord{0, -1}{\tilde{\mc F}}^{\textrm{nonrad}}_{ij} (u, \tilde{u}) &= \ord{1, 1}{\mc F}^{\textrm{nonrad}}_{ij} (u) + \frac{\tilde{u} - u}{2} \ord{1, 2}{\mc F}^{\textrm{nonrad}}_{ij} (u), \\
    \ord{0, 0}{\tilde{\mc F}}^{\textrm{nonrad}}_{ij} (u, \tilde{u}) &= \ord{0, 1}{\mc F}^{\textrm{nonrad}}_{ij} (u).
  \end{align}
\end{subequations}

\subsection{Expansion in spherical harmonics} \label{sec:harmonics}

It is important to note that the expressions that we derived in Sec.~\ref{sec:contributions} were for angular derivatives acting on the moments of the news tensor and not the moments of the news themselves.
To obtain the moments of the news from these expressions requires us to invert these angular operators, which, in turn, requires us to determine the conditions under which these angular operators can be inverted.
This issue can be addressed by expanding these moments in tensor spherical harmonics.
As we will now show, these operators \emph{cannot} be inverted for generic moments of the news greater than the second, because the operator that arises in Eq~\eqref{eqn:Nn3_bondi}, for the order $n + 3$ moment of the news annihilates harmonics with $\ell < n + 3$.
This issue does not arise for zeroth, first, and second moments of the news, however.
To show this quantitatively, we first need to introduce our conventions for our tensor spherical harmonics.

\subsubsection{Tensorial spherical harmonics}

We first introduce the tensorial spherical harmonics, which are defined as follows:\footnote{Note that the definitions of these harmonics here agree with the three types of harmonics used in~\cite{Nichols2017}, with the exception of the $s = 1$ and $I = B$ case, which differs by a minus sign.}
\begin{equation} \label{eqn:harmonic_def}
  \begin{split}
    (T^I_{\ell m})_{i_1 \cdots i_s} &\equiv 2^{(s - 1)/2} \sqrt{\frac{(\ell - s)!}{(\ell + s)!}} \\
    &\hspace{1em}\times \STF \begin{cases}
      \ms D_{i_1} \cdots \ms D_{i_s} Y_{\ell m} & I = E \\
      \epsilon_{ji_1} \ms D_{i_2} \cdots \ms D_{i_s} \ms D^j Y_{\ell m} & I = B
    \end{cases}.
  \end{split}
\end{equation}
The value $I = E$ refers to the ``electric'' harmonics, and $I = B$ the ``magnetic'' harmonics.
The index $\ell$ must satisfy $\ell \geq s$, and $m$ is in the range $-\ell \leq m \leq \ell$.\footnote{We place the tensor indices outside of parentheses so that they are not confused with the labels $I$ and $\ell m$ for the harmonics.}
From these definitions, one can show the following ``raising'' relation between rank $s - 1$ and rank $s$ harmonics:
\begin{equation} \label{eqn:E_raise}
  \STF \ms D_{i_1} (T^I_{\ell m})_{i_2 \cdots i_s} = \sqrt{\frac{(\ell - s + 1) (\ell + s)}{2}} (T^I_{\ell m})_{i_1 \cdots i_s}.
\end{equation}
We will also need the ``lowering'' relations of the form
\begin{equation} \label{eqn:E_lower_2}
  \ms D^j (T^I_{\ell m})_{ij} = -\sqrt{\frac{(\ell + 2) (\ell - 1)}{2}} (T^I_{\ell m})_i
\end{equation}
and
\begin{equation} \label{eqn:E_lower_1}
  \ms D^i (T^I_{\ell m})_i = \begin{dcases}
    -\sqrt{\ell (\ell + 1)} Y_{\ell m} & I = E \\
    0 & I = B
  \end{dcases}
\end{equation}
which are also straightforward to show.
Finally, the action of the Laplacian $\ms D^2$ on the tensor spherical harmonics is given by
\begin{align} \label{eqn:D2_harmonic}
  \ms D^2 (T^I_{\ell m})_{ij} = [4 - \ell (\ell + 1)] (T^I_{\ell m})_{ij}
\end{align}
(see, for example,~\cite{Sandberg1978}).

\subsubsection{Multipolar decomposition}

We now write the moments of the news in an expansion in tensor harmonics (where, for simplicity, we drop the dependence on $u_1$ and $u_0$):
\begin{equation}
  \ord{n}{\mc N}_{ij} \equiv \sum_{\ell \geq 2} \sum_{|m| \leq \ell} \sum_{I = E, B} \ord{n}{\mc N}_{\ell m}^I (T^I_{\ell m})_{ij}.
\end{equation}
We then consider the angular operators that act on the moments of the news in Eqs.~\eqref{eqn:N0_bondi}, \eqref{eqn:N1_bondi}, \eqref{eqn:N2_bondi}, and~\eqref{eqn:Nn3_bondi}.

First, we focus on the angular operator in Eq.~\eqref{eqn:N0_bondi}.
Equations~\eqref{eqn:E_lower_2} and~\eqref{eqn:E_lower_1} imply that
\begin{equation} \label{eqn:E_lower2_2}
  \ms D_i \ms D_j (T^I_{\ell m})^{ij} = \begin{dcases}
    \sqrt{\frac{(\ell + 2) (\ell + 1) \ell (\ell - 1)}{2}} Y_{\ell m} & I = E \\
    0 & I = B
  \end{dcases},
\end{equation}
so that
\begin{equation}
  \ms D_i \ms D_j \ord{0}{\mc N}^{ij} = \sum_{\ell \geq 2} \sum_{|m| \leq \ell}  \sqrt{\frac{(\ell + 2) (\ell + 1) \ell (\ell - 1)}{2}} \ord{0}{\mc N}^E_{\ell m} Y_{\ell m}.
\end{equation}
Thus, Eq.~\eqref{eqn:N0_bondi} determines only the electric harmonics for the zeroth moment of the news (a well-known property of computing the displacement memory effect from the conservation equation for the mass aspect; see, for example,~\cite{Flanagan2015}).

Next, consider Eq.~\eqref{eqn:N1_bondi}, and the first moment of the news.
After commuting several partial derivatives and using the definition of the Riemann tensor in two dimensions, we find
\begin{equation}
  \begin{split}
    \ms D_k \STF &(\ms D_i \ms D_j) (T^I_{\ell m})^{jk} \\
    &= \frac{1}{2} \left[2 \ms D_i \ms D_j \ms D_k (T^I_{\ell m})^{jk} - \ms D^j (\ms D^2 - 4) (T^I_{\ell m})_{ij}\right].
  \end{split}
\end{equation}
Combining this expression with those in Eqs.~\eqref{eqn:harmonic_def}, ~\eqref{eqn:D2_harmonic}, and~\eqref{eqn:E_lower2_2}, we arrive at the result
\begin{equation} \label{eqn:harmonicN1}
  \begin{split}
    \ms D_k \STF (\ms D_i \ms D_j) \ord{n}{\mc N}^{jk} & \\
    = \sum_{\ell \geq 2} \sum_{|m| \leq \ell} &\frac{\ell (\ell + 1)}{2} \sqrt{\frac{(\ell + 2) (\ell - 1)}{2}} \\
    \times &\left[\ord{n}{\mc N}_{\ell m}^E (T^E_{\ell m})_i - \ord{n}{\mc N}_{\ell m}^B (T^B_{\ell m})_i\right].
  \end{split}
\end{equation}
For $n = 1$, this is the angular operator acting on the first moment of the news, which appears in Eq.~\eqref{eqn:N1_bondi}.
It is possible to invert this equation to solve for the full first moment of the news, including both the electric and magnetic parts.

Applying a divergence to Eq.~\eqref{eqn:harmonicN1} and taking the STF part, then with the help of Eq.~\eqref{eqn:E_raise}, we obtain
\begin{equation}
  \begin{split}
    \STF \ms D_i \bigg[\ms D_l \STF (\ms D_j \ms D_k) &\ord{n}{\mc N}^{kl}\bigg] \\
    = \sum_{\ell \geq 2} \sum_{|m| \leq \ell} &\frac{(\ell - 1) \ell (\ell + 1) (\ell + 2)}{4} \\
    \times &\left[\ord{n}{\mc N}_{\ell m}^E (T^E_{\ell m})_{ij} - \ord{n}{\mc N}_{\ell m}^B (T^B_{\ell m})_{ij}\right].
  \end{split}
\end{equation}
Thus also when $n = 2$, one can invert Eq.~\eqref{eqn:N2_bondi} to obtain the all the harmonics of the second moment of the news (and hence the complete moment).

Finally, we consider Eq.~\eqref{eqn:Nn3_bondi}, and the order $n + 3$ moment of the news.
Here, we need to use Eqs.~\eqref{eqn:D_n_def} and~\eqref{eqn:D2_harmonic} to see that
\begin{equation}
  \mc D_n (T^I_{\ell m})_{ij} = \frac{(n + 2) \left[\ell (\ell + 1) - (n + 2) (n + 3)\right]}{2(n + 1) (n + 4)} (T^I_{\ell m})_{ij}.
\end{equation}
It follows that $\mc D_n$ will only annihilate the $\ell = n + 2$ harmonic.
The operator $\mc D_n \cdots \mc D_0$ will then only annihilate all harmonics with $\ell \leq n + 2$.
This implies that
\begin{widetext}
  \begin{equation}
    \begin{split}
      \mc D_n \cdots \mc D_0 \STF \ms D_i \left[\ms D_l \STF (\ms D_j \ms D_k) \ord{n + 3}{\mc N}^{kl}\right] = \sum_{\ell \geq n + 3} \sum_{|m| \leq \ell} &\left\{\prod_{k = 0}^n \frac{(k + 2) \left[\ell (\ell + 1) - (k + 2) (k + 3)\right]}{2(k + 1) (k + 4)}\right\} \\
      \times &\frac{(\ell - 1) \ell (\ell + 1) (\ell + 2)}{4} \left[\ord{n + 3}{\mc N}_{\ell m}^E (T^E_{\ell m})_{ij} - \ord{n + 3}{\mc N}_{\ell m}^B (T^B_{\ell m})_{ij}\right].
    \end{split}
  \end{equation}
\end{widetext}
One can invert this equation for the harmonics of $\ord{n + 3}{\mc N}_{ij}$ with $\ell \geq n + 3$.

In the linear theory, it follows from Eq.~\eqref{eqn:dot_En} and the fact that $\mc D_n$ annihilates the $\ell = n + 2$ harmonic that the $\ell = n + 2$ harmonic of $\ord{n + 1}{\mc E}_{ij}$ is constant for $n \geq 0$.
Newman and Penrose~\cite{Newman1968} derived an analogous result using the subleading components of the Weyl tensor instead of the components of the metric used here [though they are related by Eq.~\eqref{eqn:Riemann_E} in the linear theory].
In particular, in the case where $n = 0$, these constants are related to the \emph{Newman-Penrose constants}~\cite{Newman1968}, which are defined in terms of the subleading part of a particular component of the Weyl tensor.
The Newman-Penrose constants, unlike the $\ell = 2$ parts of $\ord{1}{\mc E}_{ij}$, are constant not only in the linear theory, but nonlinearly as well.
Whether such constants arise in the full nonlinear theory for any further subleading parts of the Weyl tensor is still an open question.

\section{Discussion} \label{sec:discussion}

In this paper, we have computed the curve deviation observable defined in Paper I~\cite{Flanagan2019a} in vacuum, asymptotically flat spacetimes near null infinity (that is, at leading order in $1/r$).
This observable generalizes geodesic deviation in that it consists of a part of the final displacement of nearby observers who have an initial displacement, relative velocity, relative acceleration, and higher nonzero time derivatives of the relative acceleration.
The dependence of the curve deviation on the initial displacement contains the displacement memory effect, the dependence on initial relative velocity contains the spin and center-of-mass memory effects, and the dependence on the initial acceleration and its time derivatives contains a number of new, independent persistent observables.
All these different observables can be cast as different ``moments of the news tensor'': namely, integrals of the product of the news tensor and an integer power of the retarded time.

Similarly to the case of the displacement memory effect, we find that the contributions to all parts of the curve deviation observable can be divided into ``charge'' and ``flux'' contributions.
The charge part corresponds to a change in a quantity that is constant in the absence of radiation, and the flux part is an integral of a quantity that vanishes in the absence of radiation.
This is perhaps not surprising given the close relation between the displacement, spin, and center-of-mass memory effects on the one hand, and the first two moments of the news that enter into the curve deviation observable.
However, this classification holds for all parts of the curve deviation and all moments of the news.
The form of the flux contribution is more involved for the dependence of the curve deviation observable on initial relative acceleration (and its higher derivatives), which constrain the second and higher moments of the news tensor.
Specifically, we find that the flux contribution can be split into two parts: (i) there is a radiative flux contribution that is given entirely in terms of quantities that appear in the gravitational waveform and (ii) a nonradiative flux contribution, which depends on source properties that are not present in the radiation.
We also found that there are moments of the news tensor in these observables that are not constrained through these charge and flux contributions, as determining these observables involves inverting angular operators that annihilate certain spherical harmonics.

In future work, we plan to compute these observables for astrophysical sources, such as compact binary mergers.
In this context, we plan to both explore these observables in a post-Newtonian framework, as has been done for the displacement memory effect~\cite{Wiseman1991}, for example, or for the spin and center-of-mass memory effects~\cite{Nichols2017, Nichols2018}.
We will also investigate these moments of the news tensor for waveforms produced by numerical relativity simulations of binary black hole systems.

Once we have predictions from astrophysical sources, we intend to assess how these effects could be measured by gravitational wave detectors such as LIGO and LISA.
The main challenge is in defining what is meant by a ``measurement'' of these effects.
This arises even in the case of the usual displacement memory effect: this effect is \emph{nonlocal in time}, reflecting a change before and after a burst of gravitational waves.
Such changes are effectively zero-frequency effects, and are undetectable by detectors with sensitivity only at finite, nonzero frequencies.
What is meant by a detection of the memory, for example in~\cite{Lasky2016, Hubner2019, Boersma2020, Hubner2021}, is instead the detection of a part of the waveform which contributes to the total displacement memory effect.
This part of the waveform, which could be called the ``displacement memory signal'' (to distinguish it from the displacement memory \emph{effect}), is a function of time, and its detectability can be estimated in terms its signal-to-noise ratio.

The displacement memory signal is given by the flux contribution in Eq.~\eqref{eqn:N0_bondi}, considered not as an integral between two fixed times $u_0$ and $u_1$, but as a function of time $u$ (by integrating from some given $u_0$ until $u$).
For the first moment of the news, there is similarly a part of a waveform which contributes to the total first moment of the news.
This part of the waveform is related to the spin and center-of-mass memory signals discussed in~\cite{Nichols2017, Nichols2018}, and constructed from the flux contribution in Eq.~\eqref{eqn:N1_bondi}.
The size of the spin and center-of-mass memory signals is even smaller than that of the displacement memory signal; thus, Refs.~\cite{Nichols2017, Nichols2018} provided preliminary evidence that the effects would not be detected until the next generation of gravitational-wave detectors following LIGO and Virgo are built.
It is natural to consider generalizing this procedure for higher moments of the news.
This is a topic we intend to pursue in future work.

\acknowledgments

We thank Keefe Mitman for helpful discussions.
D.A.N.\ acknowledges support from the NSF Grant No.\ PHY-2011784.
Many calculations in this paper were performed using the computer algebra system \textsc{Maxima}~\cite{Maxima}.
Finally, we thank the authors of~\cite{Blanchet2023} for pointing out typos in our original versions of Eqs.~\eqref{eqn:10_G_rad} and~\eqref{eqn:11_G_rad}, which we have now corrected.

\appendix

\section{Hypersurface and evolution equations} \label{app:hyper}

In this appendix, we list the relevant hypersurface and evolution equations that arise from the vacuum Einstein equations, as mentioned in Sec.~\ref{sec:coordinates}.

The first is the hypersurface equation arising from $R_{rr} = 0$, which gives the following differential equation for $\beta$:
\begin{equation} \label{eqn:hyper_beta}
  \partial_r \left(\frac{\beta}{r}\right) = \frac{r}{16} (\mc I^{-1})^{ijkl} (\partial_r \mc H_{ij}) (\partial_r \mc H_{kl}),
\end{equation}
where
\begin{equation}
  \mc I_{ijkl} \equiv \mc H_{ik} \mc H_{jl}, \qquad (\mc I^{-1})^{ijkl} \equiv (\mc H^{-1})^{ik} (\mc H^{-1})^{jl}.
\end{equation}
This differential equation can be readily solved for $\beta$, and the solution shows that $\beta$ will have no logarithmic dependence on $r$ and that $\tilde{\beta} \equiv r \beta$ is finite as $r \to \infty$.
This confirms Eq.~\eqref{eqn:beta_exp}.
Moreover, this differential equation has no constant of integration, as $\beta$ is assumed to be finite as $r \to \infty$.

The second set comes from $R_{ri} = 0$, and is a differential equation for $\mc U^i$:
\begin{equation} \label{eqn:hyper_U}
  \partial_r \left[r^4 e^{-2 \beta/r} \mc H_{ij} \partial_r \left(\frac{\mc U^j}{r^2}\right)\right] = 2 \left[r^4 \partial_r \left(\frac{\ms D_i \beta}{r^3}\right) - \mc Q_i\right],
\end{equation}
where
\begin{equation}
  \mc Q_i \equiv \frac{r^2}{2} \left\{\ms D_k [(\mc H^{-1})^{jk} \partial_r \mc H_{ij}] - \frac{1}{2} (\mc H^{-1})^{jk} \ms D_i \partial_r \mc H_{jk}\right\},
\end{equation}
which is $O(1)$ since $\partial_r \mc H_{ij} = O(1/r^2)$.
Equation~\eqref{eqn:hyper_U} can be integrated twice in order to solve for $\mc U^i$; in order to avoid logarithmic terms in the first integral, the integrand must not have an $O(1/r)$ piece.
This can be shown to imply that $\ms D^j \mc C_{ij}$ must not have a contribution at $O(1/r)$, and since divergence-free, trace-free, rank two tensors on the sphere must vanish, it follows that $\mc C_{ij}$ cannot have a contribution at $O(1/r)$, yielding Eq.~\eqref{eqn:C_exp}.
This first integral, moreover, has a constant of integration that is directly related to the angular momentum aspect $N_i$.
The second integral has neither a constant of integration (since $\mc U^i$ must remain finite as $r \to \infty$), nor any logarithmic terms.
All of these results confirm Eq.~\eqref{eqn:U_exp}.

The final two sets of Einstein equations come from $R_{ij} = 0$, which reduce to the following differential equations for $\partial_u \mc H_{ij}$ and $V$:
\begin{equation} \label{eqn:hyper_dHdu_V}
  r \mc D_{ij}{}^{kl} \left(r \partial_u \mc H_{kl} + \frac{1}{r} \mc K_{kl}\right) + 2 (\partial_r V) \mc H_{ij} = \frac{1}{r} \mc Q_{ij},
\end{equation}
where the differential operator $\mc D_{ij}{}^{kl}$ is defined by
\begin{equation}
    \mc D_{ij}{}^{kl} A_{kl} \equiv \partial_r A_{(ij)} + A_{k(i} \mc H_{j)l} \partial_r (\mc H^{-1})^{kl},
\end{equation}
the quantity $\mc K_{ij}$ defined by
\begin{equation} \label{eqn:calK}
  \mc K_{ij} \equiv \ms D_i (\mc H_{jk} \mc U^k) - \mc U^k \mc H_{kl} \mc H^l{}_{ij} - \frac{1}{2} r^2 \left(1 - \frac{2V}{r}\right) \partial_r \mc H_{ij},
\end{equation}
and the source $\mc Q_{ij}$ takes the form
\begin{widetext}
\begin{equation}
  \begin{split}
    \mc Q_{ij} \equiv r e^{2 \beta/r} \Bigg[&e^{-2 \beta/r} \mc H_{ij} - h_{ij} - \frac{1}{r} \left(e^{-2 \beta/r} \left\{\mc K_{(ij)} + \frac{1}{2r} \ms D_k [\mc U^k \partial_r (r^2 \mc H_{ij})]\right\} - 2 \ms D_i \ms D_j \beta + r \ms H_{ij}\right) \\
    &+ \frac{2}{r^2} \bigg\{(\ms D_i \beta) (\ms D_j \beta) + r \mc H^k{}_{l(i|} \left[\frac{1}{2r} e^{-2 \beta/r} \mc U^l \partial_r (r^2 \mc H_{|j)k}) - \delta^l{}_{|j)} \ms D_k \beta\right] \\
    &\hspace{3em}+ \frac{1}{4} r^6 e^{-4 \beta/r} \mc I_{ijkl} \left[\partial_r \left(\frac{\mc U^k}{r^2}\right)\right] \left[\partial_r \left(\frac{\mc U^l}{r^2}\right)\right]\bigg\}\Bigg].
  \end{split}
\end{equation}
\end{widetext}
In these equations, the quantities $\mc H^i{}_{jk}$ and $\ms H_{ij}$ are defined by
\begin{equation}
  \mc H^i{}_{jk} \equiv \frac{1}{2} (\mc H^{-1})^{il} (2 \ms D_{(j} \mc H_{k)l} - \ms D_l \mc H_{jk})
\end{equation}
and
\begin{equation}
  \ms H_{ij} \equiv \ms D_k \mc H^k{}_{ij} - \mc H^k{}_{li} \mc H^l{}_{kj}.
\end{equation}

Equation~\eqref{eqn:hyper_dHdu_V} is more complicated than either of the previous hypersurface equations, but it can be split into its pure trace and $\STF$ pieces, using $\mc H_{ij}$ instead of $h_{ij}$ (as we have done for the rest of the paper).
Contracting Eq.~\eqref{eqn:hyper_dHdu_V} into $(\mc H^{-1})^{ij}$, we find that the pure trace equation is given by
\begin{equation} \label{eqn:hyper_V}
  \partial_r V = \frac{1}{4} \left\{\frac{1}{r} (\mc H^{-1})^{ij} \mc Q_{ij} - r \partial_r \left[\frac{1}{r} (\mc H^{-1})^{ij} \mc K_{ij}\right]\right\}.
\end{equation}
Since $(\mc H^{-1})^{ij} \mc K_{ij}$ and $(\mc H^{-1})^{ij} \mc Q_{ij}$ are independent of $V$, this equation can be readily integrated to determine $V$.
The constant of integration that appears is the mass aspect $m$, and it can be shown that no logarithmic terms appear in the solution, confirming Eq.~\eqref{eqn:V_exp}.

Taking the $\STF_{\mc H}$ of Eq.~\eqref{eqn:hyper_dHdu_V} (the symmetric, trace-free part computed using $\mc H_{ij}$), we find a differential equation that only involves $\partial_u \mc H_{ij}$:
\begin{equation} \label{eqn:hyper_dHdu}
  \mc D_{ij}{}^{kl} \left(r \partial_u \mc H_{kl} + \frac{1}{r} \STF_{\mc H} \mc K_{kl}\right) = \frac{1}{r^2} \mc P_{ij},
\end{equation}
where
\begin{equation} \label{eqn:calP}
  \mc P_{ij} \equiv \STF_{\mc H} \mc Q_{ij} - \frac{r}{2} \epsilon^{kl} \mc K_{kl} \epsilon_{m(i} \mc H_{j)n} \partial_r (\mc H^{-1})^{mn}.
\end{equation}
This equation is typically called an evolution equation (by, for example,~\cite{Madler2016}), instead of a hypersurface equation.
However, it resembles a hypersurface equation, in the sense that it is a differential equation in $r$ for some new quantity $\partial_u \mc H_{ij}$, but it is a much more complicated differential equation that cannot be solved simply by an integration.
Instead, we introduce two quantities, $\mc J^{ij}{}_{kl}$ and $(\mc J^{-1})^{ij}{}_{kl}$, where the former solves the differential equation
\begin{equation}
  \partial_r \mc J^{ij}{}_{kl} = -\frac{1}{2} (\mc I^{-1})^{ij}{}_{mn} (\partial_r \mc I^{mn}{}_{op}) \mc J^{op}{}_{kl}
\end{equation}
[with the boundary condition that $\mc J^{ij}{}_{kl} = \delta^i{}_k \delta^j{}_l + O(1/r)$], and the second is an inverse in the sense that
\begin{equation}
  (\mc J^{-1})^{ij}{}_{mn} \mc J^{mn}{}_{kl} = \mc J^{ij}{}_{mn} (\mc J^{-1})^{mn}{}_{kl} = \delta^i{}_k \delta^j{}_l.
\end{equation}
In terms of these quantities, Eq.~\eqref{eqn:hyper_dHdu} can be reduced to an ordinary differential equation in $r$, which yields (after some further manipulations) the following integral expression for $\partial_u \mc C_{ij}$:
\begin{equation} \label{eqn:dot_C}
  \begin{split}
    \partial_u \mc C_{ij} &= \STF_h \left[\left(N_{kl} + \int \frac{\ud r}{r^2} \mc P_{mn} \mc J^{mn}{}_{kl}\right) (\mc J^{-1})^{kl}{}_{ij}\right] \\
    &\hspace{1em}- \frac{1}{r} \left[\STF_h \mc K_{ij} - \frac{1}{2r} \mc C_{ij} (\mc H^{-1})^{kl} \mc K_{kl}\right].
  \end{split}
\end{equation}
The constant of integration $N_{ij}$ in the first line of this expression is the news tensor, and it can be seen that this integral has no logarithmic contributions.
Expanding this equation order by order in $1/r$, one can determine the evolution equations for each $\ord{n}{\mc E}_{ij}$, such as Eqs.~\eqref{eqn:dot_E0} and~\eqref{eqn:dot_E1} for $\partial_u \ord{0}{\mc E}_{ij}$ and $\partial_u \ord{1}{\mc E}_{ij}$.

Moreover, one can show that linearizing Eq.~\eqref{eqn:dot_C} yields Eq.~\eqref{eqn:dot_En}.
To do so, we first note that Eq.~\eqref{eqn:hyper_beta} implies that
\begin{equation}
  \beta \simeq 0,
\end{equation}
and so Eq.~\eqref{eqn:hyper_U} can be integrated to obtain
\begin{equation} \label{eqn:hyper_U_lin}
  \mc U^i \simeq -\frac{2}{3r} N^i - r^2 \int \frac{\ud r}{r^4} \int r^2 \ud r \partial_r \left(\frac{\ms D_j \mc C^{ij}}{r}\right),
\end{equation}
where we have also used the linearization of Eq.~\eqref{eqn:U_1}.
Next, one can show from Eq.~\eqref{eqn:calK} that
\begin{equation}
  \mc K_{ij} \simeq \ms D_i \mc U_j - \frac{r^2}{2} \partial_r \left(\frac{\mc C_{ij}}{r}\right),
\end{equation}
and so from Eq.~\eqref{eqn:calP} that
\begin{equation}
  \mc P_{ij} \simeq \STF \mc Q_{ij} \simeq -\STF \ms D_i \mc U_j + \frac{r^2}{2} \partial_r \left(\frac{\mc C_{ij}}{r}\right),
\end{equation}
where we have used the fact that
\begin{equation}
  \STF \ms D_k \ms D_i \mc C_j{}^k = \frac{1}{2} (\ms D^2 + 2) \mc C_{ij}.
\end{equation}
As such, Eq.~\eqref{eqn:dot_C} becomes
\begin{equation}
  \partial_u \mc C_{ij} \simeq N_{ij} - \int \frac{\ud r}{r} \partial_r \STF \ms D_i \mc U_j + \frac{1}{2} \partial_r \mc C_{ij},
\end{equation}
which [together with
\begin{equation}
  \STF \ms D_i \ms D^k \mc E_{jk} = \frac{1}{2} (\ms D^2 - 2) \mc E_{ij}
\end{equation}
and Eq.~\eqref{eqn:hyper_U_lin}] implies that
\begin{widetext}
\begin{equation} \label{eqn:dot_E_lin}
  \partial_u \mc E_{ij} \simeq \frac{1}{3} \STF \ms D_i N_j + \frac{r^2}{2} \left[\int \frac{\ud r}{r} \partial_r \left\{r^2 \int \frac{\ud r}{r^4} \int r^2 \ud r \partial_r \left[\frac{(\ms D^2 - 2) \mc E_{ij}}{r^3}\right]\right\} + \partial_r \left(\frac{\mc E_{ij}}{r^2}\right)\right].
\end{equation}
\end{widetext}
Expanding this equation order-by-order in $1/r$ gives Eq.~\eqref{eqn:dot_En}.

\section{Christoffel symbols} \label{app:Christoffel}

In this appendix, we provide the Christoffel symbols of the metric in Bondi-Sachs form given in Eq.~\eqref{eqn:metric}.
A full list of these Christoffel symbols can be found (for example) in~\cite{Bondi1962} for the axisymmetric case, or~\cite{Barnich2010} for the general case, but for brevity we only list the orders in $1/r$, as it is only this information which is relevant for the discussion in this paper.
We find that
\begin{subequations}
  \begin{align}
    \Gamma^u{}_{uu} &= O(1/r), &\Gamma^r{}_{uu} &= O(1/r), &\Gamma^i{}_{uu} &= O(1/r^2), \\
    \Gamma^u{}_{ur} &= 0, &\Gamma^r{}_{ur} &= O(1/r^2), &\Gamma^i{}_{ur} &= O(1/r^3), \\
    \Gamma^u{}_{ui} &= O(1/r), &\Gamma^r{}_{ui} &= O(1/r), &\Gamma^i{}_{uj} &= O(1/r), \\
    \Gamma^u{}_{rr} &= 0, &\Gamma^r{}_{rr} &= O(1/r^2), &\Gamma^i{}_{rr} &= 0, \\
    \Gamma^u{}_{ri} &= 0, &\Gamma^r{}_{ri} &= O(1/r), &\Gamma^i{}_{rj} &= O(1/r), \\
    \Gamma^u{}_{ij} &= O(r), &\Gamma^r{}_{ij} &= O(r), &\Gamma^i{}_{jk} &= O(1).
  \end{align}
\end{subequations}
Using these equations in the geodesic equation and the equations of parallel transport, one can recover Eqs.~\eqref{eqn:dot_chi_v} and~\eqref{eqn:dyad_prop}, respectively.

\section{Derivatives of relative acceleration} \label{app:rel_acc}

In this appendix, we provide a proof of Eq.~\eqref{eqn:rel_acc}, which relates derivatives of the relative acceleration $a^a$ to derivatives of the acceleration $\ddot{\bar \gamma}^{\bar a}$ of the worldline $\bar{\gamma}$.
To do so, we first note that, for arbitrary $\tau_1 > \tau_0$,
\begin{equation} \label{eqn:acc_holonomy}
  \begin{split}
    a^{a'} &= g^{a'}{}_{\bar a'} \ddot{\bar \gamma}^{\bar a'} \\
    &= \pb{\gamma} g^{a'}{}_a g^a{}_{\bar a} \pb{\bar \gamma} g^{\bar a}{}_{\bar a'} (\Lambda^{-1})^{\bar a'}{}_{\bar b'} (\bar{\gamma}, \gamma; \tau_0) \ddot{\bar \gamma}^{\bar b'},
  \end{split}
\end{equation}
where $(\Lambda^{-1})^{\bar a'}{}_{\bar b'} (\bar{\gamma}, \gamma; \tau_0)$ is the holonomy that corresponds to parallel transport around the loop
\begin{equation}
  \bar{\gamma} (\tau_1) \to \gamma(\tau_1) \to \gamma(\tau_0) \to \bar{\gamma} (\tau_0) \to \bar{\gamma} (\tau_1)
\end{equation}
(the specific notation for this holonomy matches that of Paper I).
From the discussion in Paper I, the difference between this holonomy and the identity is a correction of $O(\bs \xi, \dot{\bs \xi}, \ddot{\bar{\bs \gamma}})$, which we can neglect as it is being multiplied by $\ddot{\bar \gamma}^{\bar a}$ in Eq.~\eqref{eqn:acc_holonomy}.
It follows that
\begin{equation} \label{eqn:acc_props}
  a^{a'} = \pb{\gamma} g^{a'}{}_a g^a{}_{\bar a} \pb{\bar \gamma} g^{\bar a}{}_{\bar a'} \ddot{\bar \gamma}^{\bar a'} + O(\bs \xi, \dot{\bs \xi}, \ddot{\bar{\bs \gamma}})^2.
\end{equation}
The definition of the parallel propagator is such that
\begin{equation}
  \frac{\uD}{\ud \tau_1} \pb{\gamma} g^{a'}{}_a = 0, \qquad \frac{\uD}{\ud \tau_1} \pb{\bar \gamma} g^{\bar a}{}_{\bar a'} = 0,
\end{equation}
which implies that, taking the $n$th derivative of Eq.~\eqref{eqn:acc_props} with respect to $\tau_1$,
\begin{equation}
  \frac{\uD^n a^{a'}}{\ud \tau_1^n} = \pb{\gamma} g^{a'}{}_a g^a{}_{\bar a} \pb{\bar \gamma} g^{\bar a}{}_{\bar a'} \frac{\uD^n \ddot{\bar \gamma}^{\bar a'}}{\ud \tau_1^n} + O(\bs \xi, \dot{\bs \xi}, \ddot{\bar{\bs \gamma}})^2.
\end{equation}
Taking the limit $\tau_1 \to \tau_0$, we recover Eq.~\eqref{eqn:rel_acc}.

\section{Integrations by parts} \label{app:ibp}

In this appendix, we prove Eqs.~\eqref{eqn:DeltaH_moment} and~\eqref{eqn:Deltaalpha_moment}, showing exactly the manipulations that are used in deriving them from Eqs.~\eqref{eqn:DeltaH_Riemann} and~\eqref{eqn:Deltaalpha_Riemann}, respectively.
This is mostly performed by integrations by parts.

Consider an arbitrary function $f(u)$.
We start with the following expression, for arbitrary $u_0 \leq u_2 \leq u_1$:
\begin{equation} \label{eqn:simple_ibp}
  \begin{split}
    \int_{u_2}^{u_1} \ud u_3 &\int_{u_2}^{u_3} \ud u_4 (u_4 - u_2) \dot{f} (u_4) \\
    &= \int_{u_2}^{u_1} \ud u_3 \bigg[(u_3 - u_2) f(u_3) - \int_{u_2}^{u_3} \ud u_4 f(u_4)\bigg],
  \end{split}
\end{equation}
which follows from an integration by parts.
Moreover, we have that
\begin{equation}
  \begin{split}
    (u_1 - u_3) f(u_3) &= \frac{\ud}{\ud u_3} \left[(u_1 - u_3) \int_u^{u_3} \ud u_4 f(u_4)\right] \\
    &\hspace{1em}+ \int_u^{u_3} \ud u_4 f(u_4),
  \end{split}
\end{equation}
for any $u$.
Setting $u = u_2$, we find that the term in brackets vanishes when $u_3 = u_2$ and when $u_3 = u_1$; thus, we have that
\begin{equation}
  \int_{u_2}^{u_1} \ud u_3 \int_{u_2}^{u_3} \ud u_4 f(u_4) = \int_{u_2}^{u_1} \ud u_3 (u_1 - u_3) f(u_3).
\end{equation}
Combining this with Eq.~\eqref{eqn:simple_ibp}, we find that
\begin{equation} \label{eqn:remove_dot}
  \begin{split}
    \int_{u_2}^{u_1} \ud u_3 &\int_{u_2}^{u_3} \ud u_4 (u_4 - u_2) \dot{f} (u_4) \\
    &= \int_{u_2}^{u_1} \ud u_3 [(u_3 - u_2) - (u_1 - u_3)] f(u_3) \\
    &= 2 \int_{u_2}^{u_1} \ud u_3 (u_3 - u_0) f(u_3) \\
    &\hspace{1em}- [(u_2 - u_0) + (u_1 - u_0)] \int_{u_2}^{u_1} \ud u_3 f(u_3).
  \end{split}
\end{equation}

We now apply Eq.~\eqref{eqn:remove_dot} to Eq.~\eqref{eqn:DeltaH_Riemann}, noting that the integral in that equation takes the form of Eq.~\eqref{eqn:remove_dot}, with $u_2 = u_0$.
Here, the function $f$ is given by the news $N^i{}_j$, and so we find
\begin{equation}
  \begin{split}
    \int_{u_0}^{u_1} \ud u_2 &\int_{u_0}^{u_2} \ud u_3 (u_3 - u_0) \dot{N}^i{}_j (u_4) \\
    &= 2 \ord{1}{\mc N}^i{}_j (u_1, u_0) - (u_1 - u_0) \ord{0}{\mc N}^i{}_j (u_1, u_0).
  \end{split}
\end{equation}
Using this equation, together with Eqs.~\eqref{eqn:DeltaH_Riemann} and~\eqref{eqn:Riemann}, we find Eq.~\eqref{eqn:DeltaH_moment}.

In order to prove Eq.~\eqref{eqn:Deltaalpha_moment} from Eq.~\eqref{eqn:Deltaalpha_Riemann}, we note that the integral on the left-hand side of Eq.~\eqref{eqn:remove_dot} appears in Eq.~\eqref{eqn:Deltaalpha_Riemann}, but multiplied by a factor of $(u_2 - u_0)^n$ and integrated from $u_0$ to $u_1$.
We then use the fact that
\begin{equation} \label{eqn:int_reduce}
  \begin{split}
    (u_2 - u_0)^n &\int_u^{u_2} \ud u_3 f(u_3) \\
    &= \frac{\ud}{\ud u_2} \left[\frac{(u_2 - u_0)^{n + 1}}{n + 1} \int_u^{u_2} \ud u_3 f(u_3)\right] \\
    &\hspace{1em}- \frac{(u_2 - u_0)^{n + 1}}{n + 1} f(u_2),
  \end{split}
\end{equation}
for any $u$; applying this equation to the case $u = u_1$ and noting (once again) that the term in brackets on the right-hand side vanishes when $u_2 = u_0$ or $u_2 = u_1$, we find that
\begin{equation}
  \begin{split}
    \int_{u_0}^{u_1} \ud u_2 (u_2 - u_0)^n &\int_{u_2}^{u_1} \ud u_3 f(u_3) \\
    &= \frac{1}{n + 1} \int_{u_0}^{u_1} \ud u_2 (u_2 - u_0)^{n + 1} f(u_2).
  \end{split}
\end{equation}
We can then combine several expressions to find that
\begin{widetext}
\begin{equation}
  \begin{split}
    \frac{1}{n!} \int_{u_0}^{u_1} \ud u_2 (u_2 - u_0)^n \int_{u_2}^{u_1} \ud u_3 \int_{u_2}^{u_3} \ud u_4 (u_4 - u_2) \dot{N}^i{}_j (u_4) &= \frac{2}{n!} \int_{u_0}^{u_1} \ud u_2 (u_2 - u_0)^n \int_{u_2}^{u_1} \ud u_3 (u_3 - u_0) N^i{}_j (u_3) \\
    &\hspace{1em}- \frac{1}{n!} \int_{u_0}^{u_1} \ud u_2 (u_2 - u_0)^{n + 1} \int_{u_2}^{u_1} \ud u_3 N^i{}_j (u_3) \\
    &\hspace{1em}- \frac{1}{n!} (u_1 - u_0) \int_{u_0}^{u_1} \ud u_2 (u_2 - u_0)^n \int_{u_2}^{u_1} \ud u_3 N^i{}_j (u_3) \\
    &= (n + 3) \ord{n + 2}{\mc N}^i{}_j (u_1, u_0) - (u_1 - u_0) \ord{n + 1}{\mc N}^i{}_j (u_1, u_0).
  \end{split}
\end{equation}
\end{widetext}
The first equality follows directly from using Eq.~\eqref{eqn:remove_dot}, and the second follows by applying Eq.~\eqref{eqn:int_reduce} to each of the three terms on the right-hand side of the first line, using $f(u) = N^i{}_j (u)$ in the second and third terms, and $f(u) = (u - u_0) N^i{}_j (u)$ in the first term.
Using this equation, together with Eqs.~\eqref{eqn:Deltaalpha_Riemann} and~\eqref{eqn:Riemann}, recovers Eq.~\eqref{eqn:Deltaalpha_moment}.

\bibliography{refs}

\end{document}